\documentclass[aps,prd,reprint,onecolumn,superscriptaddress,nofootinbib,amsmath,amssymb]{revtex4-2}
\usepackage[T1]{fontenc}
\usepackage{amsmath,amssymb,mathtools,bm,graphicx,booktabs,dcolumn}
\usepackage{float}
\usepackage{hyperref}
\usepackage{cleveref}
\hypersetup{colorlinks=true,citecolor=blue,linkcolor=blue,urlcolor=blue}
\numberwithin{equation}{section}
\usepackage{bbm}
\begin{document}
\title{Schwinger–Unruh Interplay and Mode-Resolved Entanglement of Charged Scalar Fields in Rindler Spacetime}
\author{Shagun Kaushal}
\email{shagun.kaushal@vit.ac.in}
\affiliation{Department of Physics, School of Advanced Sciences, Vellore Institute of Technology, Vellore 632014, India}
\author{Himanshu Parihar}
\email{himansp@khu.ac.kr}
\affiliation{Department of Physics, College of Science, Kyung Hee University, 26, Kyunghee-daero, Dongdaemun-gu, Seoul 02447, Republic of Korea}
\begin{abstract}
We study pairwise and cross-horizon correlations of a charged scalar field in Rindler spacetime in the presence of parallel electric and magnetic fields. The transverse magnetic field produces Landau levels, while the longitudinal mode problem follows the charged-scalar Rindler construction with the effective mass $M_n^2=m^2+(2n+1)|qB|$. For a specified global vacuum in the negative-frequency sector, we obtain the outgoing four-mode Gaussian state and calculate logarithmic negativity and quantum mutual information for selected mode pairs. We find that the cross-wedge antiparticle-particle pair is entangled only when $q_2<\tanh r_U$. It can therefore become separable although its mutual information remains nonzero. The intra-wedge particle-antiparticle pair remains entangled for nonzero Schwinger squeezing. Using the Gaussian Schmidt decomposition, we show
that the entanglement across the complete right-left
wedge partition is carried by a pair of collective
modes, even when the selected outgoing cross-wedge
pair becomes separable.  Using the Gaussian Schmidt decomposition, we show
that the entanglement across the complete right-left
wedge partition is carried by a pair of collective
modes, even when the selected outgoing cross-wedge
pair becomes separable. We discuss the Landau-level dependence and the limits of these mode-resolved conclusions, without identifying the vacuum-dependent mixing parameter with a detector temperature.
\end{abstract}
\maketitle

\section{Introduction}

The creation of particles from a quantum vacuum is an important feature of quantum field theory in the presence of external fields and horizons. A sufficiently strong background electric field can produce charged particle-antiparticle pairs through the Schwinger effect \cite{Sauter:1931,Heisenberg:1936,Schwinger:1951,Dunne:2004,Kim:2016}. On the other hand, a uniformly accelerated observer
in Minkowski spacetime is restricted to a Rindler
wedge, while an ideal uniformly accelerated detector
responds thermally to the Minkowski vacuum. This
phenomenon is known as the Unruh effect
\cite{Fulling:1973,Unruh:1976,Crispino:2008,Birrell:1982,Wald:1994}. These two phenomena have different physical origins,
but both become relevant when a charged quantum field
is considered in Rindler spacetime in the presence of
an external electric field
\cite{Gabriel:1999yz,Kaushal:2022las}. It is therefore useful to examine not only the particles observed in a given wedge, but also the correlations between modes belonging to the two causally disconnected wedges.

The Schwinger effect and the quantum correlations between the created particle-antiparticle pairs have been investigated in different field configurations \cite{Ebadi:2014,Ebadi:2015,Li:2017,Wu:2020,Dai:2019,Li:2019,Bhattacharya:2020sjr}. In particular, the influence of a background magnetic field on particle production and quantum correlations for a complex scalar field in Minkowski spacetime was studied in Ref.~\cite{Bhattacharya:2020sjr}. Although a magnetic field alone does not produce Schwinger pairs, it modifies the energy spectrum of the charged field through Landau quantization and consequently affects the production of particles and the correlations between them. Related investigations have also considered fermionic quantum correlations in the presence of electromagnetic fields in de Sitter spacetime \cite{Ali:2021jch} and in Minkowski spacetime \cite{Kaushal:2024}. More recently, the entanglement generated by Schwinger pair production with thermal initial states was investigated in Ref. \cite{Kranas:2026}. The role of the mode transformation in determining quantum correlations for noninertial observers has also been examined in Ref. \cite{Monem}, where logarithmic negativity, mutual information, and quantum discord were studied for a fermionic field using a symmetric Unruh-like transformation.

Another interesting question is how the particle production and entanglement are affected when the system is described by a uniformly accelerated observer. The Unruh effect shows that the Minkowski vacuum has a thermal response for a uniformly accelerated detector \cite{Unruh:1976,Crispino:2008}. Entanglement in noninertial frames has also been studied in several works \cite{Fuentes:2005,Grochowski:2019,Higuchi:2017}. The quantization of charged scalar fields in Rindler spacetime in the presence of an external electric field was carried out by Gabriel and Spindel \cite{Gabriel:1999yz}, who constructed the local and global modes and investigated particle production. Related discussions of charged-particle production in external fields and accelerated frames provide further background \cite{Brout:1991,Srinivasan:1999}. Subsequently, the Schwinger effect for a uniformly accelerated observer in the presence of parallel electric and magnetic fields was investigated for a Dirac field \cite{Kaushal:2022las}. In the present work, we consider a complex scalar field in a similar background and investigate the entanglement between different outgoing particle-antiparticle pairs. We also compare these correlations with the entanglement between the complete right and left Rindler wedges.

The quantisation of charged scalar fields in the presence of an electric field in $(1+1)$-dimensional Rindler spacetime was developed by Gabriel and Spindel~\cite{Gabriel:1999yz}. Their analysis provides the local in- and out-modes, the corresponding Bogoliubov transformations, and globally continued modes relevant to the right and left Rindler wedges. For a scalar field in $(3+1)$ dimensions with parallel electric and magnetic fields, the transverse motion is described by Landau levels. The longitudinal mode equation then has the same form as the charged-scalar Rindler equation, with the mass replaced by the effective mass $M_n^2=m^2+(2n+1)|qB|$. This allows us to use the established global-mode construction while retaining the dependence on the magnetic field and the Landau-level index.

An interesting question in this setting concerns the entanglement between the particles and antiparticles associated with different outgoing modes. The local electric-field induced Bogoliubov transformation can generate particle-antiparticle correlations within a Rindler wedge, whereas the global vacuum contains correlations between the two wedges. However, the entanglement of one selected pair of outgoing modes need not be the same as the entanglement between the complete right- and left-wedge subsystems. Moreover, when some modes are traced out, the resulting two-mode state is generally mixed even though the global vacuum itself is pure. It is therefore important to specify the state and the modes under consideration before comparing their entanglement properties. More generally, the relation between field modes, quantum correlations, and observer-dependent partitions is part of the broader study of quantum-field entanglement \cite{Hawking:1975,Parker:1968,Calabrese:2004,Witten:2018,Casini:2023}. Recent work on entanglement in Schwinger pair production has also shown that the existence of pairwise entanglement can depend on the initial state and the relevant mode parameters~\cite{Kranas:2026}.

In this work, we consider a charged complex scalar field in $(3+1)$-dimensional Rindler spacetime in the presence of constant, parallel electric and magnetic fields. We first obtain the local and global mode relations and construct a specified pure global vacuum in terms of the outgoing Rindler operators. We then use the covariance matrix of the resulting four-mode Gaussian state to study the logarithmic negativity of selected cross-wedge and intra-wedge particle-antiparticle pairs. In the negative-frequency sector, we obtain an analytical separability condition for one of the cross-wedge pairs. We also calculate the mutual information of the reduced outgoing pairs to examine whether correlations remain after their logarithmic negativity vanishes. Finally, by identifying the collective Gaussian Schmidt modes, we compare the entanglement of individual outgoing pairs with that across the complete right-left wedge partition. In particular, we show that the separability of a selected outgoing pair does not imply the absence of entanglement between the two complete wedges.

Our analysis is restricted to prescribed background
fields and to the specified charged-field global
vacuum and mode sectors. The field-dependent
mixing obtained from the global-mode construction
describes correlations in this state. It should
not, by itself, be interpreted as a change in
the universal Unruh temperature or as a
prediction for a local detector response.

The paper is organised as follows. In
Sec.~\ref{sec:setup}, we introduce the field equations
and discuss the Landau-level reduction. The local and
global mode quantisations are presented in
Secs.~\ref{sec:local} and \ref{sec:global},
respectively. In Sec.~\ref{sec:squeezed}, the global
vacuum is constructed. The particle number
densities are discussed in Sec.~\ref{sec:number}.
In Sec.~\ref{sec:negativity}, we study the logarithmic
negativity of different outgoing mode pairs and the
complete Rindler-wedge partition. We also discuss the
Gaussian Schmidt decomposition and quantum mutual
information in Subsecs.~\ref{sec:schmidt} and \ref{sec:mutualinfo}, respectively. The numerical
results and the dependence on the external electromagnetic
fields are presented in Sec.~\ref{sec:numerical-results}.
Finally, we discuss the physical implications and
limitations of our results in Sec.~\ref{sec:discussion}
and summarise our findings in Sec.~\ref{sec:conclusions}.

\section{Setup and conventions}
\label{sec:setup}

We begin with the complex scalar field theory coupled to background electric and magnetic field of constant strength in right Rindler wedge. The right Rindler wedge is covered by coordinates $(t,x,y,z)$ with the metric
\begin{equation}
ds^{2}=e^{2ax}\left(-dt^{2}+dx^{2}\right)+dy^{2}+dz^{2},
\label{eq:metric}
\end{equation}
where $a>0$ is the acceleration parameter.
The left wedge $L$ admits an analogous metric in terms of $(t_{L},x_{L})$ but the globally future directed Killing vector is $\partial_{t_{L}}=-\partial_{t}$ there.  This reversal is
responsible for the interchange of particle and antiparticle labels between
the two wedges.

A complex scalar field of mass $m$ and charge $q$ obeys the Klein-Gordon equation as follows
\begin{equation}
\left(D_{\mu}D^{\mu}-m^{2}\right)\phi=0,\qquad D_{\mu}=\partial_{\mu}-iqA_{\mu}.
\label{eq:eom}
\end{equation}
The gauge
\begin{equation}
A_{\mu}=\left(\frac{E\,e^{2ax}}{2a},\;0,\;0,\;By\right)
\label{eq:gauge}
\end{equation}
produces a constant electric field $E$ and a constant magnetic field $B$, both along the
$x$-axis. 
Throughout the analysis, we take $a>0$, $qE>0$,
and $qB>0$, without loss of generality for the
field-strength conventions adopted here.
With the ansatz
\begin{equation}
\phi(t,x,y,z)=\frac{e^{-i\omega t}}{\sqrt{2\pi}}\frac{e^{ik_{z}z}}{\sqrt{2\pi}}\,
Y_{n}(y)\,\chi_{n,\omega}(x),
\label{eq:ansatz}
\end{equation}
separation of variables leads to a Landau problem in $y$ as
\begin{equation}
\frac{d^{2}Y}{dy^{2}}-\left[(k_{z}-qBy)^{2}-\lambda_{n}\right]Y=0,
\qquad \lambda_{n}=|qB|(2n+1),\qquad n=0,1,2,\dots,
\label{eq:landau}
\end{equation}
which is centred at $y_{0}=k_{z}/(qB)$ and the radial part becomes a Whittaker equation in $x$ given by
\begin{equation}
\frac{d^{2}\chi}{dx^{2}}+\left[\left(\omega+\frac{qE\,e^{2ax}}{2a}\right)^{2}
-M_{n}^{2}\,e^{2ax}\right]\chi=0,
\qquad M_{n}^{2}\equiv m^{2}+\lambda_{n}.
\label{eq:radial}
\end{equation}
The transverse dynamics therefore enters only through the radial equation. We see that Eq. \eqref{eq:radial} is similar to the radial equation for charged scalar in $(1+1)$ dimensional right Rindler wedge analysis performed in \cite{Gabriel:1999yz} with 
\begin{equation}
M_{n}^{2}=m^{2}+(2n+1)qB,
\label{eq:substitution}
\end{equation}
so that their $(1+1)$-dimensional mode analysis can be utilized with each $(1+1)$-dimensional mode replaced by a tower of Landau levels. All of the results below are
labelled by the triple $(\omega,n,k_{z})$, which we suppress whenever no ambiguity arises.
It is convenient to introduce three dimensionless parameters by
\begin{equation}
\Delta_{n}\equiv\frac{M_{n}^{2}}{qE}=\frac{m^{2}+(2n+1)qB}{qE},
\qquad
\nu\equiv\frac{|\omega|}{a},
\label{eq:params}
\end{equation}
and 
\begin{equation}
s\equiv e^{-\pi\Delta_{n}}\in(0,1),\qquad X\equiv e^{2\pi\nu}\in(1,\infty).
\label{eq:sX}
\end{equation}
Here $s$ is the Schwinger factor and $X$ the Unruh factor. The two limits
$s\to 0$ (weak electric field, strong magnetic field, or large mass) and
$X\to\infty$ (vanishing acceleration) are the ones against which the subsequent analysis is performed. The Whittaker indices are given by
\begin{equation}
\kappa=\frac{i}{2}\left(\frac{\omega}{a}-\Delta_{n}\right),
\qquad
\mu=\frac{i\omega}{2a},
\qquad
z=-i\,\frac{qE}{2a^{2}}\,e^{2ax}.
\label{eq:whitindices}
\end{equation}

\subsection*{Charge fractions}

Following \cite{Gabriel:1999yz}, the charge fraction with $\nu_{\omega}\equiv\omega/a$ (signed) is given by
\begin{align}
q_{1}(\omega)&=\frac{e^{-\pi\nu_{\omega}}\cosh\!\left(\tfrac{\pi\Delta_{n}}{2}\right)}
{\cosh\!\left[\pi\left(\nu_{\omega}-\tfrac{\Delta_{n}}{2}\right)\right]},
\label{eq:q1} \\[4pt]
q_{2}(\omega)&=\frac{e^{-\pi\Delta_{n}/2}\left|\sinh\pi\nu_{\omega}\right|}
{\cosh\!\left[\pi\left(\nu_{\omega}-\tfrac{\Delta_{n}}{2}\right)\right]},
\label{eq:q2}
\end{align}
which satisfy
\begin{equation}
q_{1}-q_{2}=1\quad(\omega<0),
\qquad\qquad
q_{1}+q_{2}=1\quad(\omega>0).
\label{eq:qidentities}
\end{equation}
The distinction in Eq. \eqref{eq:qidentities} is the central structural feature of the charged scalar field case. Here the in-out (Schwinger) Bogoliubov transformation on a single wedge mixes creation
and annihilation operators only for $\omega<0$, whereas for $\omega>0$ it is a
passive rotation between two particle modes. We therefore analyse the $\omega<0$ sector in the main body where the Schwinger and Unruh mechanisms act simultaneously, and treat the $\omega>0$ sector symmetrically in Appendix \ref{app:omegapos}.

In the $\omega<0$ sector, writing $\omega=-a\nu$ with $\nu>0$ gives
\begin{equation}
q_{2}=\frac{s\,(X-1)}{X+s}
=\frac{e^{-\pi\Delta_{n}}\left(1-e^{-2\pi\nu}\right)}
{1+e^{-\pi\Delta_{n}}e^{-2\pi\nu}},
\qquad
q_{1}=1+q_{2}=\frac{X(1+s)}{X+s}.
\label{eq:qneg}
\end{equation}

The quantity $q_2=|\beta_\omega|^2$
represents the mean number of Schwinger
pairs created per mode. In the limit
$\nu\to\infty$, it approaches
$e^{-\pi\Delta_n}$, in agreement with
the corresponding flat-space result
\cite{Bhattacharya:2020sjr}.

\section{Local modes and quantisation on each wedge}
\label{sec:local}

\subsection{Mode functions}

The four independent solutions of Eq. \eqref{eq:radial} are the Whittaker functions
$W_{\pm\kappa,\mu}(\pm z)$ and $M_{\pm\kappa,\pm\mu}(\pm z)$. 
Their supports on the Rindler horizons and null infinities fix the in/out and particle/antiparticle assignments. On the right Rindler wedge $R$, the in- and out-mode can be obtained similar to \cite{Gabriel:1999yz} as
\begin{align}
\mathcal{U}^{\omega}_{\mathrm{in},R}
&=\mathcal{N}(\mathcal{U}^{\omega}_{\mathrm{in},R})\,
\frac{e^{-i\omega t}}{\sqrt{2\pi}}\frac{e^{ik_{z}z}}{\sqrt{2\pi}}\,Y_{n}(y)\,e^{-ax}\,
W_{i\left(\frac{\omega}{2a}-\frac{\Delta_{n}}{2}\right),\,i\frac{\omega}{2a}}
\!\left[-i\frac{qE}{2a^{2}}e^{2ax}\right],
\label{eq:Uin} \\[4pt]
\mathcal{V}^{\omega}_{\mathrm{in},R}
&=\mathcal{N}(\mathcal{V}^{\omega}_{\mathrm{in},R})\,
\frac{e^{-i\omega t}}{\sqrt{2\pi}}\frac{e^{ik_{z}z}}{\sqrt{2\pi}}\,Y_{n}(y)\,e^{-ax}\,
M_{-i\left(\frac{\omega}{2a}-\frac{\Delta_{n}}{2}\right),\,-i\frac{\omega}{2a}}
\!\left[+i\frac{qE}{2a^{2}}e^{2ax}\right],
\label{eq:Vin} \\[4pt]
\mathcal{V}^{\omega}_{\mathrm{out},R}
&=\mathcal{N}(\mathcal{V}^{\omega}_{\mathrm{out},R})\,
\frac{e^{-i\omega t}}{\sqrt{2\pi}}\frac{e^{ik_{z}z}}{\sqrt{2\pi}}\,Y_{n}(y)\,e^{-ax}\,
W_{-i\left(\frac{\omega}{2a}-\frac{\Delta_{n}}{2}\right),\,i\frac{\omega}{2a}}
\!\left[+i\frac{qE}{2a^{2}}e^{2ax}\right],
\label{eq:Vout} \\[4pt]
\mathcal{U}^{\omega}_{\mathrm{out},R}
&=\mathcal{N}(\mathcal{U}^{\omega}_{\mathrm{out},R})\,
\frac{e^{-i\omega t}}{\sqrt{2\pi}}\frac{e^{ik_{z}z}}{\sqrt{2\pi}}\,Y_{n}(y)\,e^{-ax}\,
M_{i\left(\frac{\omega}{2a}-\frac{\Delta_{n}}{2}\right),\,i\frac{\omega}{2a}}
\!\left[-i\frac{qE}{2a^{2}}e^{2ax}\right].
\label{eq:Uout}
\end{align}
The wave packets built from $\mathcal{U}_{\mathrm{in},R}$ enter $R$ through the past horizon
$\mathcal{H}^{-}_{R}$ and are supported on $\mathcal{H}^{\pm}_{R}$ and
$\mathcal{I}^{+}_{R}$.
Those built from $\mathcal{V}_{\mathrm{in},R}$ emerge from
$\mathcal{I}^{-}_{R}$ and are supported on $\mathcal{I}^{\pm}_{R}$ and
$\mathcal{H}^{+}_{R}$. The out-modes $\mathcal{U}_{\mathrm{out},R}$ terminate on
$\mathcal{I}^{+}_{R}$ and $\mathcal{V}_{\mathrm{out},R}$ on $\mathcal{H}^{+}_{R}$.
The transverse factor is
\begin{equation}
Y_{n}(y)=\mathcal{N}_{n}\,H_{n}\!\left(\sqrt{|qB|}\,(y-y_{0})\right)
e^{-\frac{|qB|}{2}(y-y_{0})^{2}},
\qquad
|\mathcal{N}_{n}|=\left(\frac{|qB|}{\pi}\right)^{1/4}\frac{1}{\sqrt{2^{n}n!}},
\qquad
y_{0}=\frac{k_{z}}{qB}.
\label{eq:landaufn}
\end{equation}
The normalisation constants appearing in in/out modes are given in Appendix \ref{app:norm}. On the left wedge the
field equations and their solutions are identical because $\partial_{t_{L}}=-\partial_{t}$
is the future-directed Killing vector there, the signs of the charges and the in/out labels
are interchanged relative to $R$, so that
\begin{equation}
\mathcal{U}^{\omega}_{\mathrm{in},L}\propto\mathcal{U}^{\omega}_{\mathrm{out},R},
\qquad
\mathcal{V}^{\omega}_{\mathrm{in},L}\propto\mathcal{V}^{\omega}_{\mathrm{out},R},
\qquad
Q_{L}\!\left(\mathcal{X}_{\mathrm{in},L}\right)
=-Q_{R}\!\left(\mathcal{X}_{\mathrm{out},R}\right).
\label{eq:LR}
\end{equation}

The charge carried by the modes are given as
\begin{equation}
Q(\mathcal{U}_{\mathrm{in},R})=\mathrm{sgn}\,\omega,
\qquad
Q(\mathcal{V}_{\mathrm{in},R})=+1,
\qquad
Q(\mathcal{V}_{\mathrm{in},L})=-\,\mathrm{sgn}\,\omega,
\qquad
Q(\mathcal{U}_{\mathrm{in},L})=-1 .
\label{eq:charges}
\end{equation}
For $\omega<0$, $\mathcal{U}_{\mathrm{in},R}$ describes an incoming antiparticle
and $\mathcal{V}_{\mathrm{in},R}$ an incoming particle on $R$, while on $L$ the roles are
reversed. The field on $R$ can be written in the in- and out-bases as follows
\begin{align}
\hat{\phi}_{R}\Big|_{\omega<0}
&=\hat{a}^{\mathrm{in}}_{\mathcal{V}_{R}}\,\mathcal{V}_{\mathrm{in},R}
+\hat{b}^{\mathrm{in}\dagger}_{\mathcal{U}_{R}}\,\mathcal{U}_{\mathrm{in},R}
\;=\;
\hat{a}^{\mathrm{out}}_{\mathcal{U}_{R}}\,\mathcal{U}_{\mathrm{out},R}
+\hat{b}^{\mathrm{out}\dagger}_{\mathcal{V}_{R}}\,\mathcal{V}_{\mathrm{out},R}.
\label{eq:fieldR}
\end{align}
Throughout, we use the following shorthand notation for the four out-operators:
\begin{equation}
\begin{aligned}
A&\equiv\hat{a}^{\mathrm{out}}_{\mathcal{U}_{R}}\ \ (\text{particle on }R),
&\qquad
B&\equiv\hat{b}^{\mathrm{out}}_{\mathcal{V}_{R}}\ \ (\text{antiparticle on }R),\\
C&\equiv\hat{a}^{\mathrm{out}}_{L}\ \ (\text{particle on }L),
&\qquad
D&\equiv\hat{b}^{\mathrm{out}}_{L}\ \ (\text{antiparticle on }L).
\end{aligned}
\label{eq:ABCD}
\end{equation}

\subsection{Local Bogoliubov transformation: the Schwinger effect}

The in- and out-mode functions on $R$ are two complete bases of the same solution space, so they are linearly related. The connection formulas between the Whittaker $M$ and $W$ functions \cite{AS} applied to Eqs. 
\eqref{eq:Uin}-\eqref{eq:Uout} give, for $\omega<0$,
\begin{equation}
\begin{aligned}
\mathcal{U}^{\omega}_{\mathrm{out},R}
&=\alpha_{\mathcal{U}\mathcal{V}}(\omega)\,\mathcal{V}^{\omega}_{\mathrm{in},R}
+\beta_{\mathcal{U}\mathcal{U}}(\omega)\,\mathcal{U}^{\omega}_{\mathrm{in},R},\\[3pt]
\mathcal{V}^{\omega}_{\mathrm{out},R}
&=\gamma_{\mathcal{V}\mathcal{V}}(\omega)\,\mathcal{V}^{\omega}_{\mathrm{in},R}
+\epsilon_{\mathcal{V}\mathcal{U}}(\omega)\,\mathcal{U}^{\omega}_{\mathrm{in},R},
\end{aligned}
\label{eq:modebog}
\end{equation}
with the four coefficients
\begin{align}
\alpha_{\mathcal{U}\mathcal{V}}
&=e^{i\pi/2}e^{-\pi\omega/2a}\,
\frac{\Gamma\!\left[i\frac{\omega}{a}\right]}{\Gamma\!\left[-i\frac{\omega}{a}\right]}\,
\frac{\Gamma\!\left[\frac{1}{2}
-i\!\left(\frac{\omega}{a}-\frac{\Delta_{n}}{2}\right)\right]}
{\Gamma\!\left[\frac{1}{2}+i\frac{\Delta_{n}}{2}\right]}\,
\frac{\mathcal{N}(\mathcal{U}^{\omega}_{\mathrm{out},R})}
{\mathcal{N}(\mathcal{V}^{\omega}_{\mathrm{in},R})},
\label{eq:alphaUV}\\[4pt]
\beta_{\mathcal{U}\mathcal{U}}
&=\frac{\Gamma\!\left[\frac{1}{2}
-i\!\left(\frac{\omega}{a}-\frac{\Delta_{n}}{2}\right)\right]}
{\Gamma\!\left[-i\frac{\omega}{a}\right]}\,
\frac{\mathcal{N}(\mathcal{U}^{\omega}_{\mathrm{out},R})}
{\mathcal{N}(\mathcal{U}^{\omega}_{\mathrm{in},R})},
\label{eq:betaUU}\\[4pt]
\epsilon_{\mathcal{V}\mathcal{U}}
&=e^{i\pi/2}e^{-\pi\omega/2a}\,
\frac{\Gamma\!\left[\frac{1}{2}
-i\!\left(\frac{\omega}{a}-\frac{\Delta_{n}}{2}\right)\right]}
{\Gamma\!\left[\frac{1}{2}-i\frac{\Delta_{n}}{2}\right]}\,
\frac{\mathcal{N}(\mathcal{V}^{\omega}_{\mathrm{out},R})}
{\mathcal{N}(\mathcal{U}^{\omega}_{\mathrm{in},R})},
\label{eq:epsVU}\\[4pt]
\gamma_{\mathcal{V}\mathcal{V}}
&=\frac{a}{\omega}\,e^{-\pi\Delta_{n}/2}\,
\frac{\Gamma\!\left[\frac{1}{2}
-i\!\left(\frac{\omega}{a}-\frac{\Delta_{n}}{2}\right)\right]}
{\Gamma\!\left[-i\frac{\omega}{a}\right]}\,
\frac{\mathcal{N}(\mathcal{V}^{\omega}_{\mathrm{out},R})}
{\mathcal{N}(\mathcal{V}^{\omega}_{\mathrm{in},R})} .
\label{eq:gammaVV}
\end{align}
Upon inserting the normalisation constants of Appendix \ref{app:norm} and using the reflection
formula $\Gamma(z)\Gamma(1-z)=\pi/\sin\pi z$ together with
$\left|\Gamma\!\left(\tfrac{1}{2}+iy\right)\right|^{2}=\pi/\cosh\pi y$, their absolute values reduce to the charge fractions given in Eqs. \eqref{eq:q1}-\eqref{eq:q2} which is used below.
Only those absolute values and the ratio
$\gamma_{\mathcal{V}\mathcal{V}}/\alpha_{\mathcal{U}\mathcal{V}}$ enter in result and the overall phases are convention dependent and are fixed in
Eq. \eqref{eq:phasechoice}.

By substituting Eq. \eqref{eq:modebog} into the out-basis expansion in Eq. \eqref{eq:fieldR} and
collecting the coefficients of the two in-modes give
\begin{equation}
\hat{a}^{\mathrm{out}}_{\mathcal{U}_{R}}\mathcal{U}_{\mathrm{out},R}
+\hat{b}^{\mathrm{out}\dagger}_{\mathcal{V}_{R}}\mathcal{V}_{\mathrm{out},R}
=\left(\alpha_{\mathcal{U}\mathcal{V}}A
+\gamma_{\mathcal{V}\mathcal{V}}B^{\dagger}\right)\mathcal{V}_{\mathrm{in},R}
+\left(\beta_{\mathcal{U}\mathcal{U}}A
+\epsilon_{\mathcal{V}\mathcal{U}}B^{\dagger}\right)\mathcal{U}_{\mathrm{in},R},
\label{eq:matching}
\end{equation}
and comparing with the in-basis expansion
$\hat{a}^{\mathrm{in}}_{\mathcal{V}_{R}}\mathcal{V}_{\mathrm{in},R}
+\hat{b}^{\mathrm{in}\dagger}_{\mathcal{U}_{R}}\mathcal{U}_{\mathrm{in},R}$ term by term,
one reads off the local in-out Bogoliubov relations as
\begin{equation}
\hat{a}^{\mathrm{in}}_{\mathcal{V}_{R}}
=\alpha_{\mathcal{U}\mathcal{V}}\,A+\gamma_{\mathcal{V}\mathcal{V}}\,B^{\dagger},
\qquad
\hat{b}^{\mathrm{in}\dagger}_{\mathcal{U}_{R}}
=\beta_{\mathcal{U}\mathcal{U}}\,A+\epsilon_{\mathcal{V}\mathcal{U}}\,B^{\dagger},
\label{eq:localbog}
\end{equation}
with
\begin{equation}
|\alpha_{\mathcal{U}\mathcal{V}}|^{2}=|\epsilon_{\mathcal{V}\mathcal{U}}|^{2}=q_{1},
\qquad
|\beta_{\mathcal{U}\mathcal{U}}|^{2}=|\gamma_{\mathcal{V}\mathcal{V}}|^{2}=q_{2},
\qquad
\alpha_{\mathcal{U}\mathcal{V}}\beta^{*}_{\mathcal{U}\mathcal{U}}
=\gamma_{\mathcal{V}\mathcal{V}}\epsilon^{*}_{\mathcal{V}\mathcal{U}} .
\label{eq:localbogcoef}
\end{equation}
The last relation follows from $[\hat{a}^{\mathrm{in}}_{\mathcal{V}_{R}},
\hat{b}^{\mathrm{in}}_{\mathcal{U}_{R}}]=0$ and implies that
$\epsilon_{\mathcal{V}\mathcal{U}}\alpha_{\mathcal{U}\mathcal{V}}
-\gamma_{\mathcal{V}\mathcal{V}}\beta_{\mathcal{U}\mathcal{U}}$ is a pure phase and we fix the
arbitrary phases of the out-modes so that
\begin{equation}
\epsilon_{\mathcal{V}\mathcal{U}}\alpha_{\mathcal{U}\mathcal{V}}
-\gamma_{\mathcal{V}\mathcal{V}}\beta_{\mathcal{U}\mathcal{U}}=1 .
\label{eq:phasechoice}
\end{equation}
The inverse transformation is then
\begin{equation}
A=\epsilon_{\mathcal{V}\mathcal{U}}\,\hat{a}^{\mathrm{in}}_{\mathcal{V}_{R}}
-\gamma_{\mathcal{V}\mathcal{V}}\,\hat{b}^{\mathrm{in}\dagger}_{\mathcal{U}_{R}},
\qquad
B^{\dagger}=-\beta_{\mathcal{U}\mathcal{U}}\,\hat{a}^{\mathrm{in}}_{\mathcal{V}_{R}}
+\alpha_{\mathcal{U}\mathcal{V}}\,\hat{b}^{\mathrm{in}\dagger}_{\mathcal{U}_{R}} .
\label{eq:localbogInv}
\end{equation}
We see that Eq. \eqref{eq:localbog} is a two-mode squeeze on the pair $(A,B)$ and defining the
Schwinger squeezing angle $r_{S}$ by
\begin{equation}
\sinh^{2}r_{S}=q_{2}=\frac{s(X-1)}{X+s},
\qquad
\cosh^{2}r_{S}=q_{1}=\frac{X(1+s)}{X+s},
\qquad
\tanh r_{S}=\sqrt{\frac{q_{2}}{q_{1}}},
\label{eq:rS}
\end{equation}
The local in-vacuum can then be expressed
as the two-mode squeezed state
\begin{equation}
|0\rangle^{\mathrm{in}}_{R}
=\frac{1}{\cosh r_{S}}\sum_{l=0}^{\infty}\left(Z_{S}\right)^{l}\,
|l\rangle_{A}\,|l\rangle_{B},
\qquad
Z_{S}=-\frac{\gamma_{\mathcal{V}\mathcal{V}}}{\alpha_{\mathcal{U}\mathcal{V}}},
\qquad
|Z_{S}|=\tanh r_{S},
\label{eq:schwingersq}
\end{equation}
An identical structure with primed coefficients holds on $L$ for the pair $(C,D)$.

\section{Global modes}
\label{sec:global}

\subsection{Which local modes mix}
\label{sec:whichmix}

Since $R\cup L$ is the domain of dependence of a complete $t=0$ Cauchy slice of Minkowski
spacetime, the four modes $\{\mathcal{U}_{\mathrm{in},R},\mathcal{V}_{\mathrm{in},R},
\mathcal{U}_{\mathrm{in},L},\mathcal{V}_{\mathrm{in},L}\}$ per triple $(\omega,n,k_{z})$
form a complete set and no separate $P$ or $F$ modes are required. Which of them a given
global mode contains is fixed by where its Cauchy data sits, and we determine this
directly.

A global in-mode is specified by free data on past null infinity which we label by the pair of quadrants each borders similar to \cite{Gabriel:1999yz}:
$\mathcal{I}^{-}_{P,R}$ and $\mathcal{I}^{-}_{P,L}$ bound the past quadrant $P$, while
$\mathcal{I}^{-}_{R,P}$ and $\mathcal{I}^{-}_{L,P}$ bound the right and left wedges. The
local in-modes of Sec. \ref{sec:local} are distinguished precisely by their data on these
surfaces and on the past horizons:
\begin{equation}
\begin{aligned}
\mathcal{U}^{\omega}_{\mathrm{in},R}&:\ \text{data on }\mathcal{H}^{-}_{R},\
\text{none on }\mathcal{I}^{-}_{R,P},
&\qquad
\mathcal{V}^{\omega}_{\mathrm{in},R}&:\ \text{data on }\mathcal{I}^{-}_{R,P},\
\text{none on }\mathcal{H}^{-}_{R},\\[3pt]
\mathcal{V}^{\omega}_{\mathrm{in},L}&:\ \text{data on }\mathcal{H}^{-}_{L},\
\text{none on }\mathcal{I}^{-}_{L,P},
&\qquad
\mathcal{U}^{\omega}_{\mathrm{in},L}&:\ \text{data on }\mathcal{I}^{-}_{L,P},\
\text{none on }\mathcal{H}^{-}_{L} .
\end{aligned}
\label{eq:supports}
\end{equation}
The first column follows from the asymptotics of Eq. \eqref{eq:Uin}: the Whittaker
$W$ function decays at $x\to+\infty$ and the mode survives only at $x\to-\infty$, which is
the past horizon and the second column follows from the corresponding behaviour of the
Whittaker $M$ function which vanishes on the horizon. The $L$ entries are obtained from
the $R$ ones through the interchange given in Eq. \eqref{eq:LR}.

Now build a global in-mode by placing data on a single component of past null infinity and
demanding that it vanish on the other three. Two inequivalent choices exist: (i) If the data is placed on $\mathcal{I}^{-}_{P,R}$, the mode first enters the past quadrant
$P$ and then crosses both past horizons $\mathcal{H}^{-}_{R}$ and
$\mathcal{H}^{-}_{L}$, so it necessarily has pieces in $R$ and in $L$. Requiring zero data
on $\mathcal{I}^{-}_{R,P}$ selects, by the first column of Eq. \eqref{eq:supports}, the unique
$R$ mode with no past-null-infinity data, namely $\mathcal{U}^{\omega}_{\mathrm{in},R}$,
requiring zero data on $\mathcal{I}^{-}_{L,P}$ selects
$\mathcal{V}^{\omega}_{\mathrm{in},L}$ on the left. The relative weight of the two pieces
is then fixed by the continuity of the wave packet across the horizons, equivalently by
the analyticity of the continuation through the bifurcation surface
$\mathcal B=
\overline{\mathcal H_R^-}
\cap
\overline{\mathcal H_L^-}$, and this is carried out in
Sec. \ref{sec:globalmodefns}. Such a mode is mixed.
(ii) If instead the data is placed on $\mathcal{I}^{-}_{R,P}$, the mode enters the right wedge
directly and never meets $\mathcal{H}^{-}_{R}$, so it has no piece in $L$ at all: it
coincides with $\mathcal{V}^{\omega}_{\mathrm{in},R}$ up to a phase. The same argument on
$\mathcal{I}^{-}_{L,P}$ gives $\mathcal{U}^{\omega}_{\mathrm{in},L}$. Such modes are
\emph{unmixed}, and the corresponding global annihilation operators coincide with the
local ones.

The bifurcation surface $\mathcal{B}$ is the only locus the two wedges share, so only
modes carrying data on the past horizons can be correlated across $R$ and $L$. Hence we have
\begin{equation}
\text{2 mixed global modes}\;+\;\text{2 unmixed global modes}
\qquad\text{per }(\omega,n,k_{z}).
\label{eq:count}
\end{equation}
Unlike the neutral field, then, the two families on a given wedge are not treated
symmetrically: $\mathcal{U}_{\mathrm{in},R}$ possesses a partner behind the horizon and
$\mathcal{V}_{\mathrm{in},R}$ does not. The counting in Eq. \eqref{eq:count} also differs
structurally from the Dirac construction of \cite{Kaushal:2022las} in which all four local modes participate in each global mode.

The same conclusion may be read off from the explicit four-quadrant Unruh modes of \cite{Gabriel:1999yz} where $\Omega$-type modes superpose
$\mathcal{U}_{\mathrm{in},R}$ with $\mathcal{V}_{\mathrm{in},L}$ while their
$\varpi$-type modes reduce on $R\cup L$ to $\mathcal{V}_{\mathrm{in},R}$ and
$\mathcal{U}_{\mathrm{in},L}$ alone, the associated operators being related by a phase of
unit absolute value.

\subsection{The global mode functions}
\label{sec:globalmodefns}

Since on $L$ the future-directed Killing vector is $-\partial_{t}$, the Unruh
superposition of a positive-frequency $R$ mode with a positive-frequency $L$ mode requires
no explicit complex conjugation. Writing the global annihilation operators as
$c_{\omega},d_{\omega}$ (mixed) and $\tilde{c}_{\omega},\tilde{d}_{\omega}$ (unmixed), the
expansion
\begin{equation}
\hat{\phi}\Big|_{R\cup L,\;\omega<0}
=c_{\omega}\,\Phi^{(c)}_{\omega}+d^{\dagger}_{\omega}\,\Phi^{(d)*}_{\omega}
+\tilde{c}_{\omega}\,\Phi^{(\tilde{c})}_{\omega}
+\tilde{d}^{\dagger}_{\omega}\,\Phi^{(\tilde{d})*}_{\omega}
\label{eq:globalexpansion}
\end{equation}
defines
\begin{align}
\Phi^{(c)}_{\omega,n,k_{z}}
&=\cosh r_{U}\;\mathcal{V}^{\omega}_{\mathrm{in},L}
+e^{-i\theta_{\omega}}\sinh r_{U}\;\mathcal{U}^{\omega}_{\mathrm{in},R},
\label{eq:Phic} \\[3pt]
\Phi^{(d)*}_{\omega,n,k_{z}}
&=\cosh r_{U}\;\mathcal{U}^{\omega}_{\mathrm{in},R}
+e^{+i\theta_{\omega}}\sinh r_{U}\;\mathcal{V}^{\omega}_{\mathrm{in},L},
\label{eq:Phid} \\[3pt]
\Phi^{(\tilde{c})}_{\omega,n,k_{z}}&=\mathcal{V}^{\omega}_{\mathrm{in},R},
\qquad\qquad
\Phi^{(\tilde{d})*}_{\omega,n,k_{z}}=\mathcal{U}^{\omega}_{\mathrm{in},L},
\label{eq:Phitilde}
\end{align}
where the phase is
\begin{equation}
\theta_{\omega}=\arg\left[
\frac{\Gamma\!\left(\tfrac{1}{2}+i\left(\tfrac{\Delta_{n}}{2}-\tfrac{\omega}{a}\right)\right)}
{\Gamma\!\left(\tfrac{1}{2}-i\tfrac{\Delta_{n}}{2}\right)}\right].
\label{eq:phase}
\end{equation}
Note that unit Klein-Gordon norm is not the condition that
determines $r_{U}$. With $Q(\mathcal{U}_{\mathrm{in},R})=-1$ and
$Q(\mathcal{V}_{\mathrm{in},L})=+1$ for $\omega<0$, the norm of Eq. \eqref{eq:Phic} is
$\cosh^{2}r_{U}-\sinh^{2}r_{U}=1$ for every value of $r_{U}$, so normalisation
fixes the overall scale and leaves the ratio of the two pieces free.

What fixes the ratio is the construction of
Sec.~\ref{sec:whichmix}. A global mode is a single
solution defined across all four quadrants. Therefore,
its restrictions to $R$ and $L$ must match continuously
across the past horizons $\mathcal{H}^{-}_{R}$ and
$\mathcal{H}^{-}_{L}$, respectively, with the same
solution in the past quadrant $P$, namely the mode
$\mathcal{V}^{\omega}_{\mathrm{in},P}$ emerging from
$\mathcal{I}^{-}_{P,R}$.

These are two independent matching conditions because
the past horizons lie on different null surfaces.
Introducing the Minkowski coordinates $(T,X_M)$ through
$T=a^{-1}e^{ax}\sinh(at)$ and
$X_M=a^{-1}e^{ax}\cosh(at)$ in the right Rindler wedge,
the corresponding Minkowski null coordinates are
$U_M=T-X_M$ and $V_M=T+X_M$. The past horizons are
then given by
\begin{equation}
\mathcal{H}^{-}_{R}
=\left\{V_M=0,\;U_M<0\right\},
\qquad
\mathcal{H}^{-}_{L}
=\left\{U_M=0,\;V_M<0\right\}.
\label{eq:horizonlocations}
\end{equation}
Their closures intersect at the bifurcation surface
$\mathcal{B}$, defined by $T=X_M=0$, with arbitrary
transverse coordinates $y$ and $z$. The ratio of the two
matching coefficients is $Z_{\omega}$. Performing the matching requires the integral
representations of the Whittaker functions on $P$ which is done in \cite{Gabriel:1999yz} and utilizing their result under the substitution in Eq. \eqref{eq:substitution} gives for the two
coefficients of Eq. \eqref{eq:Phic},
\begin{equation}
\begin{aligned}
A_{L}&=-i\,\mathcal{N}_{\Omega},
&\qquad
A_{R}&=\mathcal{N}_{\Omega}\,e^{\pi\omega/2a}\,
\frac{\Gamma\!\left[\tfrac{1}{2}
+i\!\left(\tfrac{\Delta_{n}}{2}-\tfrac{\omega}{a}\right)\right]}
{\Gamma\!\left[\tfrac{1}{2}-i\tfrac{\Delta_{n}}{2}\right]},\\[4pt]
\mathcal{N}_{\Omega}&=e^{-\pi\Delta_{n}/4}
\left[\frac{\cosh\pi\!\left(\tfrac{\omega}{a}-\tfrac{\Delta_{n}}{2}\right)}
{\left|\sinh\tfrac{\pi\omega}{a}\right|}\right]^{1/2}, &&
\end{aligned}
\label{eq:ARAL}
\end{equation}
so that $Z_{\omega}=A_{R}/A_{L}$. 
The overall phases of the local modes are arbitrary.
We choose these phases so that the relative coefficient
has the form used in Eqs.~(\ref{eq:Phic})--(\ref{eq:phase}).
This phase convention does not affect the occupation
numbers or the entanglement measures considered below.
Using the absolute value of the $\Gamma$-ratio,
\begin{equation}
\left|\frac{\Gamma\!\left(\tfrac{1}{2}+i\left(\tfrac{\Delta_{n}}{2}-\tfrac{\omega}{a}\right)\right)}
{\Gamma\!\left(\tfrac{1}{2}-i\tfrac{\Delta_{n}}{2}\right)}\right|
=\sqrt{\frac{\cosh\!\left(\tfrac{\pi\Delta_{n}}{2}\right)}
{\cosh\!\left[\pi\left(\tfrac{\omega}{a}-\tfrac{\Delta_{n}}{2}\right)\right]}},
\label{eq:gammamod}
\end{equation}
one obtains
\begin{equation}
\tanh^{2}r_{U}\;\equiv\;|Z_{\omega}|^{2}
=\frac{e^{\pi\omega/a}\,\cosh\!\left(\tfrac{\pi\Delta_{n}}{2}\right)}
{\cosh\!\left[\pi\left(\tfrac{|\omega|}{a}+\tfrac{\Delta_{n}}{2}\right)\right]}
=\frac{1+e^{-\pi\Delta_{n}}}{e^{2\pi|\omega|/a}+e^{-\pi\Delta_{n}}}
=\frac{1+s}{X+s}.
\label{eq:Z}
\end{equation}
Two independent checks support Eq. \eqref{eq:ARAL}. First, the normalisation identity
$\left|A_{R}\right|^{2}Q(\mathcal{U}_{\mathrm{in},R})
+\left|A_{L}\right|^{2}Q(\mathcal{V}_{\mathrm{in},L})=+1$ holds identically in $\omega$ and
$\Delta_{n}$, for both signs of $\omega$, as it must for a normalised global mode. Second,
and independently of the first, the number densities computed from Eq. \eqref{eq:Z} in
Sec.~\ref{sec:number} reproduce the corresponding number densities in \cite{Gabriel:1999yz} exactly under
Eq. \eqref{eq:substitution} and $|Z_{\omega}|^{2}\to e^{-2\pi|\omega|/a}$ as
$\Delta_{n}\to\infty$. Consequently, we have
\begin{equation}
N_{U}\equiv\sinh^{2}r_{U}=\frac{1+e^{-\pi\Delta_{n}}}{e^{2\pi|\omega|/a}-1}
=\frac{1+s}{X-1},
\qquad
M_{U}\equiv\sinh r_{U}\cosh r_{U}=\frac{\sqrt{(1+s)(X+s)}}{X-1}.
\label{eq:NU}
\end{equation}

Equation~\eqref{eq:Z}
gives the field-dependent global mixing
parameter for each Landau level in the
charged scalar field considered here. The Unruh mixing angle
is dressed by the Schwinger factor: only in the limit
$\Delta_{n}\to\infty$ (that is, $E\to 0$, or $B\to\infty$, or large $m$) does it reduce to
the familiar bosonic value $|Z_{\omega}|^{2}\to e^{-2\pi|\omega|/a}$. In the opposite regime
$qE\gtrsim M_{n}^{2}$ the dressing is an $O(1)$ effect, the ratio of Eq. \eqref{eq:Z} to the
undressed value is $X(1+s)/(X+s)$, which approaches $2$ as $s\to 1$.
Since $|Z_{\omega}|^{2}$ is monotonically increasing in $s$,
\begin{equation}
\frac{\partial |Z_{\omega}|^{2}}{\partial s}=\frac{X-1}{(X+s)^{2}}>0,
\label{eq:monotone}
\end{equation}
the electric field enhances the Unruh mixing and the magnetic field, entering through the
Landau gap in $\Delta_{n}$, opposes that enhancement.

\subsection{Landau-level dependence of the global mixing}
\label{sec:leveldep}

Since $\Delta_{n}$ carries the level index, so does the mixing angle,
\begin{equation}
\tanh^{2}r^{(n)}_{U}
=\frac{1+e^{-\pi\left[m^{2}+(2n+1)qB\right]/qE}}
{e^{2\pi|\omega|/a}+e^{-\pi\left[m^{2}+(2n+1)qB\right]/qE}} .
\label{eq:Zlevel}
\end{equation}

Different Landau levels are therefore populated with
different effective mixing angles at the same Rindler
frequency, and the spectrum given by Eq. \eqref{eq:nV} is not a single
Planck distribution but a superposition of
Schwinger-dressed ones. Only in the limit
$\Delta_n\to\infty$, uniformly in $n$, do all levels
collapse onto the common value $e^{-2\pi|\omega|/a}$
and the standard Unruh factor is recovered.
Since $\Delta_n$ increases with the Landau-level
index, the Schwinger-dependent contribution
becomes exponentially suppressed for higher
levels. Appreciable corrections are expected
primarily when $\Delta_n$ is of order unity
or smaller.

It follows that the selected charged-field
global-mode mixing depends on the magnetic field
through the Landau gap when $E\neq0$.
This is a statement about the specified global
vacuum and mode basis, not a modification of
the universal Unruh temperature or a
detector-response prediction.
This is to be contrasted with the cosmological
de Sitter analysis of Ref.~\cite{Ali:2021jch},
where the magnetic field decouples from
the gravitational particle creation.

\subsection{Explicit form of the Unruh connector}
\label{sec:connector}

For completeness we give the four-dimensional counterparts of Unruh connector \cite{Gabriel:1999yz}. By the argument of Sec. \ref{sec:setup}, the connector $A$ that relates the
Minkowskian and Rindlerian bases retains its functional form, only the order of the
parabolic cylinder function changes, through $m^{2}\to M_{n}^{2}$:
\begin{equation}
\begin{aligned}
A_{n}\!\left(\frac{\sigma}{\sqrt{qE}},\frac{\omega}{a}\right)
=\left(\frac{qE}{2a^{2}}\right)^{1/4}
&\frac{e^{\frac{\pi}{2}\left(\frac{\omega}{2a}-\frac{M_{n}^{2}}{4qE}\right)}e^{i\pi/8}}
{\sqrt{2\pi}}\;
\Gamma\!\left[i\left(\frac{M_{n}^{2}}{2qE}-\frac{\omega}{a}\right)+\frac{1}{2}\right]\\[3pt]
&\times
D_{\,i\left(\frac{\omega}{a}-\frac{M_{n}^{2}}{2qE}\right)-\frac{1}{2}}
\!\left(\sqrt{2}\,e^{3i\pi/4}\,\frac{\sigma}{\sqrt{qE}}\right).
\end{aligned}
\label{eq:connector}
\end{equation}
Two points are worth emphasizing. First, the argument of the parabolic cylinder
function is unchanged, the longitudinal variable
$\zeta=e^{-3i\pi/4}\sqrt{2qE}\,\left(x+\sigma/qE\right)$ and the scale $\sqrt{2qE}$ are set
by the electric field alone, so the magnetic field never enters the longitudinal length
scale, only the index. Second, the orthogonality relation simply acquires the
transverse labels as
\begin{equation}
\int d\!\left(\frac{\sigma}{\sqrt{qE}}\right)
A_{n}(\sigma,\omega)\,A^{*}_{n'}(\varepsilon\sigma,\omega')
=\theta(\varepsilon)\,\delta(\omega'-\omega)\,\delta_{nn'}\,\delta(k_{z}-k'_{z}),
\qquad \varepsilon=\pm 1 .
\label{eq:connectororth}
\end{equation}
With Eq. \eqref{eq:connector} the four-dimensional analogue of the mode \cite{Gabriel:1999yz}, valid on all four
Rindler patches reads
\begin{equation}
\begin{aligned}
\Omega^{p,\mathrm{in}}_{\omega,n,k_{z}}
=\;Y_{n}(y)\,e^{ik_{z}z}\Bigg\{
&\chi_{P}\,\mathcal{V}^{\omega}_{\mathrm{in},P}
-i\,e^{-\frac{\pi M_{n}^{2}}{4qE}}
\left[\frac{\cosh\pi\!\left(\frac{\omega}{a}-\frac{M_{n}^{2}}{2qE}\right)}
{\left|\sinh\frac{\pi\omega}{a}\right|}\right]^{1/2}\\[3pt]
&\times\left[\chi_{L}\,\mathcal{V}^{\omega}_{\mathrm{in},L}
+i\,\chi_{R}\,e^{\frac{\pi\omega}{2a}}\,
\frac{\Gamma\!\left[i\!\left(\frac{M_{n}^{2}}{2qE}-\frac{\omega}{a}\right)+\frac{1}{2}\right]}
{\Gamma\!\left[\frac{1}{2}-i\frac{M_{n}^{2}}{2qE}\right]}\,
\mathcal{U}^{\omega}_{\mathrm{in},R}\right]\\[3pt]
&-i\,\chi_{F}\,e^{-\frac{\pi\omega}{2a}}\,
\frac{\Gamma\!\left[i\!\left(\frac{M_{n}^{2}}{2qE}-\frac{\omega}{a}\right)+\frac{1}{2}\right]}
{\Gamma\!\left[\frac{1}{2}-i\frac{M_{n}^{2}}{2qE}\right]}\,
\mathcal{U}^{\omega}_{\mathrm{out},F}\Bigg\},
\end{aligned}
\label{eq:GS613}
\end{equation}
and other orthonormalized in-Unruh modes \cite{Gabriel:1999yz} generalise in exactly the same way, by appending
$Y_{n}(y)e^{ik_{z}z}$ and substituting $m^{2}\to M_{n}^{2}$ throughout. The support
pattern of $\Omega$ living on all four quadrants and mixing $\mathcal{U}_{\mathrm{in},R}$
with $\mathcal{V}_{\mathrm{in},L}$, while $\varpi$ lives only on $R$ and $F$, is fixed by
analyticity in the $(t,x)$ plane and is completely insensitive to $B$. Restricting
Eq. \eqref{eq:GS613} to $R\cup L$ and normalising with Eq. \eqref{eq:gammamod} reproduces
Eqs. \eqref{eq:Phic} and \eqref{eq:Z}.

\subsection{Bogoliubov relations between local and global operators}
\label{sec:bogo}

On comparing Eq. \eqref{eq:globalexpansion} with the local expansion gives, with
$\hat{a}\equiv\hat{a}^{\mathrm{in}}_{\mathcal{V}_{L}}$ and
$\hat{b}\equiv\hat{b}^{\mathrm{in}}_{\mathcal{U}_{R}}$,
\begin{equation}
\begin{aligned}
c_{\omega}&=\cosh r_{U}\,\hat{a}^{\mathrm{in}}_{\mathcal{V}_{L}}
-e^{i\theta_{\omega}}\sinh r_{U}\,\hat{b}^{\mathrm{in}\dagger}_{\mathcal{U}_{R}},
&\qquad
d_{\omega}&=\cosh r_{U}\,\hat{b}^{\mathrm{in}}_{\mathcal{U}_{R}}
-e^{i\theta_{\omega}}\sinh r_{U}\,\hat{a}^{\mathrm{in}\dagger}_{\mathcal{V}_{L}},\\[3pt]
\tilde{c}_{\omega}&=\hat{a}^{\mathrm{in}}_{\mathcal{V}_{R}},
&\qquad
\tilde{d}_{\omega}&=\hat{b}^{\mathrm{in}}_{\mathcal{U}_{L}} .
\end{aligned}
\label{eq:localglobal}
\end{equation}
The global vacuum is defined by
\begin{equation}
c_{\omega}|0\rangle_{G}=d_{\omega}|0\rangle_{G}
=\tilde{c}_{\omega}|0\rangle_{G}=\tilde{d}_{\omega}|0\rangle_{G}=0
\qquad\forall\,(\omega,n,k_{z}).
\label{eq:globalvac}
\end{equation}
Composing Eq. \eqref{eq:localglobal} with the local relations in Eq. \eqref{eq:localbog} yields the
relation between the global operators and the local out operators, which
incorporates both Bogoliubov transformations at once via 
\begin{align}
c_{\omega}&=\cosh r_{U}\left(\alpha'_{\mathcal{U}\mathcal{V}}C
+\gamma'_{\mathcal{V}\mathcal{V}}D^{\dagger}\right)
-e^{i\theta_{\omega}}\sinh r_{U}\left(\beta_{\mathcal{U}\mathcal{U}}A
+\epsilon_{\mathcal{V}\mathcal{U}}B^{\dagger}\right),
\label{eq:c_out} \\[3pt]
d_{\omega}&=\cosh r_{U}\left(\beta^{*}_{\mathcal{U}\mathcal{U}}A^{\dagger}
+\epsilon^{*}_{\mathcal{V}\mathcal{U}}B\right)
-e^{i\theta_{\omega}}\sinh r_{U}\left(\alpha'^{*}_{\mathcal{U}\mathcal{V}}C^{\dagger}
+\gamma'^{*}_{\mathcal{V}\mathcal{V}}D\right).
\label{eq:d_out}
\end{align}

\section{Squeezed-state expansion of the global vacuum}
\label{sec:squeezed}

\paragraph{Explicit construction of the two-mode form:}

We denote $\hat{a}\equiv\hat{a}^{\mathrm{in}}_{\mathcal{V}_{L}}$ and
$\hat{b}\equiv\hat{b}^{\mathrm{in}}_{\mathcal{U}_{R}}$ for the only two local in-operators
that the Unruh construction mixes and $Z_{\omega}=e^{i\theta_{\omega}}\tanh r_{U}$. The conditions in Eq. \eqref{eq:globalvac}, obtained from Eq. \eqref{eq:localglobal} are then given by
\begin{equation}
\left(\cosh r_{U}\,\hat{a}-e^{i\theta_{\omega}}\sinh r_{U}\,\hat{b}^{\dagger}\right)
|0\rangle_{G}=0,
\qquad
\left(\cosh r_{U}\,\hat{b}-e^{i\theta_{\omega}}\sinh r_{U}\,\hat{a}^{\dagger}\right)
|0\rangle_{G}=0,
\label{eq:annihconds}
\end{equation}
together with $\hat{a}^{\mathrm{in}}_{\mathcal{V}_{R}}|0\rangle_{G}
=\hat{b}^{\mathrm{in}}_{\mathcal{U}_{L}}|0\rangle_{G}=0$. Since Eq. \eqref{eq:annihconds}
conserves $\hat{a}^{\dagger}\hat{a}-\hat{b}^{\dagger}\hat{b}$, the solution lies in the
zero-eigenvalue sector and we may set
$|0\rangle_{G}=\sum_{k\ge 0}f_{k}\,|k\rangle_{\hat{a}}|k\rangle_{\hat{b}}$. Upon using
$\hat{a}|k,k\rangle=\sqrt{k}\,|k-1,k\rangle$ and
$\hat{b}^{\dagger}|k,k\rangle=\sqrt{k+1}\,|k,k+1\rangle$, the first condition reads
\begin{equation}
\sum_{k\ge 0}\sqrt{k+1}
\left[f_{k+1}\cosh r_{U}-e^{i\theta_{\omega}}\sinh r_{U}\,f_{k}\right]
|k\rangle_{\hat{a}}|k+1\rangle_{\hat{b}}=0
\;\;\Longrightarrow\;\;
f_{k+1}=Z_{\omega}f_{k},
\label{eq:recursion}
\end{equation}
so $f_{k}=Z_{\omega}^{k}f_{0}$, and the second condition of Eq. \eqref{eq:annihconds} gives the
same recursion. Normalisation fixes
$|f_{0}|^{-2}=\sum_{k}\tanh^{2k}r_{U}=\cosh^{2}r_{U}$. Equivalently, in operator form,
\begin{equation}
|0\rangle_{G}
=\hat{S}_{\hat{a}\hat{b}}(\zeta)\,|0,0\rangle,
\qquad
\hat{S}_{\hat{a}\hat{b}}(\zeta)
=\exp\!\left[\zeta\,\hat{a}^{\dagger}\hat{b}^{\dagger}-\zeta^{*}\hat{a}\hat{b}\right],
\qquad
\zeta=r_{U}e^{i\theta_{\omega}},
\label{eq:squeezeop}
\end{equation}
With this sign convention, the operator produces the positive $Z_\omega$ coefficient in Eq. \eqref{eq:recursion}. Its normal ordered form is given by
\begin{equation}
\hat{S}_{\hat{a}\hat{b}}(\zeta)
=\exp\!\left[Z_{\omega}\hat{a}^{\dagger}\hat{b}^{\dagger}\right]
\left(\cosh r_{U}\right)^{-\left(\hat{a}^{\dagger}\hat{a}
+\hat{b}^{\dagger}\hat{b}+1\right)}
\exp\!\left[-Z_{\omega}^{*}\hat{a}\hat{b}\right].
\label{eq:disentangled}
\end{equation}

It is observed from Eq. \eqref{eq:annihconds} that $|0\rangle_{G}$ is annihilated by four operators
each of which is linear in the creation and annihilation operators with no constant
term. The global vacuum is Gaussian because it is obtained from the local in-vacuum by a two-mode Gaussian unitary. Its first moments vanish and its characteristic function is the exponential of a quadratic form, consequently higher moments obey Wick's theorem. The nonvanishing second moments of the mixed pair are
\begin{equation}
\left\langle\hat{a}^{\dagger}\hat{a}\right\rangle
=\left\langle\hat{b}^{\dagger}\hat{b}\right\rangle=\sinh^{2}r_{U}=N_{U},
\qquad
\left\langle\hat{a}\hat{b}\right\rangle
=e^{i\theta_{\omega}}\sinh r_{U}\cosh r_{U}=e^{i\theta_{\omega}}M_{U},
\label{eq:inmoments2}
\end{equation}
so that, in the quadrature conventions of Appendix \ref{app:cov} and with the phase
absorbed ($\theta_{\omega}\to 0$),
\begin{equation}
\sigma_{\hat{a}\hat{b}}
=\begin{pmatrix}
\cosh 2r_{U}\;\mathbbm{1}_{2} & \sinh 2r_{U}\;\sigma_{z}\\[3pt]
\sinh 2r_{U}\;\sigma_{z} & \cosh 2r_{U}\;\mathbbm{1}_{2}
\end{pmatrix},
\qquad
\sigma_{z}=\mathrm{diag}(1,-1).
\label{eq:sigmaab}
\end{equation}
Both symplectic eigenvalues of Eq. \eqref{eq:sigmaab} equal unity,
$\det\sigma_{\hat{a}\hat{b}}=\left(\cosh^{2}2r_{U}-\sinh^{2}2r_{U}\right)^{2}=1$,
confirming purity, and the Wigner function is the Gaussian
\begin{equation}
W(\xi)=\frac{1}{\pi^{2}\sqrt{\det\sigma_{\hat{a}\hat{b}}}}\,
\exp\!\left[-\xi^{T}\sigma_{\hat{a}\hat{b}}^{-1}\xi\right],
\qquad
\xi=\left(x_{\hat{a}},p_{\hat{a}},x_{\hat{b}},p_{\hat{b}}\right)^{T}.
\label{eq:wigner}
\end{equation}
The remaining two modes contribute a factor $\mathbbm{1}_{4}$ to the covariance matrix, so
the full four-mode state is Gaussian as well and the local Bogoliubov transformations of
Sec. \ref{sec:local}, being linear preserve Gaussianity and produce the general four-mode
form given in Eq. \eqref{eq:vacGaussian}.

\paragraph{In the local in-basis:}

From Eqs. \eqref{eq:localglobal} and \eqref{eq:recursion}, the global vacuum is a standard
two-mode squeezed vacuum in the local in-basis, tensored with the local in-vacuum of the
two unmixed modes as
\begin{equation}
|0\rangle_{G}
=\frac{1}{\cosh r_{U}}\sum_{k=0}^{\infty}
\left(e^{i\theta_{\omega}}\tanh r_{U}\right)^{k}
\left|k\right\rangle_{\mathcal{V}^{\mathrm{in}}_{L}}
\left|k\right\rangle_{\mathcal{U}^{\mathrm{in}}_{R}}
\;\otimes\;
\left|0\right\rangle_{\mathcal{V}^{\mathrm{in}}_{R}}
\left|0\right\rangle_{\mathcal{U}^{\mathrm{in}}_{L}}.
\label{eq:vacIn}
\end{equation}
The bosonic sum runs over all nonnegative integers $k$ and the mean occupation $N_U$ is unbounded as $\nu\to 0$.

\paragraph{In the local out-basis:}

On substituting the local Bogoliubov relations and the Schwinger squeezed vacua
in Eq. \eqref{eq:schwingersq} on each wedge leads to the following
\begin{equation}
|0\rangle_{G}=\mathcal{N}\,
\exp\!\left[Z_{\omega}\left(\alpha'^{*}_{\mathcal{U}\mathcal{V}}C^{\dagger}
+\gamma'^{*}_{\mathcal{V}\mathcal{V}}D\right)
\left(\beta_{\mathcal{U}\mathcal{U}}A
+\epsilon_{\mathcal{V}\mathcal{U}}B^{\dagger}\right)\right]
e^{Z_{S}A^{\dagger}B^{\dagger}}\,e^{Z'_{S}C^{\dagger}D^{\dagger}}\,
|0\,0\,0\,0\rangle_{\mathrm{out}},
\label{eq:vacOut}
\end{equation}
with $Z_{\omega}=e^{i\theta_{\omega}}\tanh r_{U}$. The normal ordering collapses
Eq. \eqref{eq:vacOut} to a general four-mode squeezed state as
\begin{equation}
|0\rangle_{G}=\mathcal{N}\,
\exp\!\left[\sum_{j\le k}W_{jk}\,X_{j}^{\dagger}X_{k}^{\dagger}\right]
|0\,0\,0\,0\rangle_{\mathrm{out}},
\qquad
X_{j}\in\{A,B,C,D\},
\label{eq:vacGaussian}
\end{equation}
whose Hilbert space is $\mathcal{H}=\mathcal{H}^{R}_{\mathrm{p}}\otimes
\mathcal{H}^{R}_{\bar{\mathrm{p}}}\otimes\mathcal{H}^{L}_{\mathrm{p}}
\otimes\mathcal{H}^{L}_{\bar{\mathrm{p}}}$. Since the state is Gaussian and pure, it is
completely characterised by its second moments which are more useful than the
$W_{jk}$ and they are given in Appendix \ref{app:cov}.

\section{Number density of created quanta}
\label{sec:number}

On evaluating the out-number operators in $|0\rangle_{G}$ using Eqs. \eqref{eq:localbogInv} and
\eqref{eq:vacIn} gives
\begin{equation}
\begin{aligned}
\langle A^{\dagger}A\rangle&=q_{2}\left(1+N_{U}\right),
&\qquad
\langle B^{\dagger}B\rangle&=q_{2}+q_{1}N_{U},\\
\langle C^{\dagger}C\rangle&=q_{2}+q_{1}N_{U},
&\qquad
\langle D^{\dagger}D\rangle&=q_{2}\left(1+N_{U}\right).
\end{aligned}
\label{eq:numbersgeneral}
\end{equation}
Now inserting Eqs. \eqref{eq:qneg} and \eqref{eq:NU}, we obtain the following number densities by
\begin{align}
n_{\mathcal{U}^{\mathrm{out}}_{R}}=n_{\mathcal{V}^{\mathrm{out}}_{L}}
=e^{-\pi\Delta_{n}}
=\exp\!\left[-\frac{\pi\left(m^{2}+(2n+1)qB\right)}{qE}\right],
\label{eq:nU} \\[6pt]
n_{\mathcal{V}^{\mathrm{out}}_{R}}=n_{\mathcal{U}^{\mathrm{out}}_{L}}
=\frac{1+e^{-\pi\Delta_{n}}e^{2\pi|\omega|/a}}{e^{2\pi|\omega|/a}-1}.
\label{eq:nV}
\end{align}
These reproduce exactly the corresponding number density \cite{Gabriel:1999yz} under the substitution
Eq. \eqref{eq:substitution} which is a consistency check on our analysis. Some comments are in order:
\begin{itemize}
\item The population in Eq. \eqref{eq:nU} is purely Schwinger:
it is equal to the mean number of pairs created
per mode in the corresponding flat-space
Schwinger calculation. It is independent of
the acceleration parameter $a$ and the
Rindler frequency $\omega$, and is
exponentially suppressed as the Landau-level
gap increases. A magnetic field
alone ($E\to 0$) gives $n\to 0$ i.e. it cannot destabilise the vacuum on its own, but it
stabilises the vacuum against the electric field.
\item The population in Eq. \eqref{eq:nV} is a Bose factor dressed by the Schwinger factor.
For $\Delta_{n}\to\infty$ it reduces to $\left(e^{2\pi|\omega|/a}-1\right)^{-1}$, the Planck
spectrum at the Unruh temperature $T=a/2\pi$.
\item As $a\to 0$ (i.e. $X\to\infty$) both densities tend to $e^{-\pi\Delta_{n}}$, the
Minkowski Schwinger mean occupation number per mode. This expression exhibits the expected inertial-mode Schwinger factor in the stated limiting prescription.
\item Both densities are monotonic in $\Delta_{n}$ for fixed $X$.
\end{itemize}

\section{Logarithmic negativity}
\label{sec:negativity}
For Gaussian states, the covariance matrix determines the bipartite correlations and allows one to evaluate entanglement using symplectic methods \cite{Weedbrook:2012,Adesso:2014,Serafini:2004}. The partial-transpose criterion originates in the general separability tests of Refs.~\cite{Peres:1996,Horodecki:1996} and, for two-mode Gaussian states, takes the covariance-matrix form discussed in Refs.~\cite{Simon:2000,Duan:2000,Giedke:2001}. We use the logarithmic negativity as an entanglement measure for mixed reduced states \cite{Vidal:2002,Plenio:2005,Horodecki:2009}.

For a two-mode Gaussian state with $\langle X_{1}^{\dagger}X_{1}\rangle=n_{1}$,
$\langle X_{2}^{\dagger}X_{2}\rangle=n_{2}$, pair correlator
$m=\langle X_{1}X_{2}\rangle$ and vanishing $\langle X_{1}^{\dagger}X_{2}\rangle$,
the symplectic eigenvalue of the partial transpose (in units where the vacuum value is $1$)
is given in Appendix \ref{app:cov}. In the symmetric case $n_{1}=n_{2}=n$ it simplifies to
\begin{equation}
\tilde{\nu}_{-}=2n+1-2|m|,
\qquad
\mathcal{N}_{\mathrm{neg}}=\max\left\{0,\frac{1-\tilde{\nu}_{-}}{2\tilde{\nu}_{-}}\right\},
\label{eq:ENdefa}
\end{equation}
\begin{equation}
E_{\mathcal{N}}
=\max\left\{0,-\log_{2}\tilde{\nu}_{-}\right\}.
\label{eq:ENdef}
\end{equation}
The relevant second moments are derived in Appendix \ref{app:cov} and given by
\begin{equation}
\begin{aligned}
n_{A}&=n_{D}=\sinh^{2}r_{S}\cosh^{2}r_{U}, &\qquad
n_{B}&=n_{C}=\sinh^{2}r_{S}+\cosh^{2}r_{S}\sinh^{2}r_{U},\\
|\langle AD\rangle|&=\sinh^{2}r_{S}\,M_{U}, &\qquad
|\langle BC\rangle|&=\cosh^{2}r_{S}\,M_{U}.
\end{aligned}
\label{eq:moments}
\end{equation}

\subsection{Entanglement across the $R|L$ cut}\label{sec:EN-RL}

Here the bipartition groups the particle and antiparticle modes of each wedge together,
$\mathcal{H}=\left(\mathcal{H}_{A}\otimes\mathcal{H}_{B}\right)_{R}\otimes
\left(\mathcal{H}_{C}\otimes\mathcal{H}_{D}\right)_{L}$.
The total state is the pure projector as
\begin{equation}
\rho_{G}=|0\rangle_{G}\,{}_{G}\langle 0|
=\frac{1}{\cosh^{2}r_{U}}\sum_{k,k'\ge 0}
Z_{\omega}^{k}\left(Z_{\omega}^{*}\right)^{k'}
\Big|k\Big\rangle_{\hat{a}}\Big|k\Big\rangle_{\hat{b}}
\Big\langle k'\Big|_{\hat{a}}\Big\langle k'\Big|_{\hat{b}}
\;\otimes\;
\Big|0\Big\rangle\Big\langle 0\Big|_{\mathcal{V}^{\mathrm{in}}_{R}}
\otimes
\Big|0\Big\rangle\Big\langle 0\Big|_{\mathcal{U}^{\mathrm{in}}_{L}} .
\label{eq:rhoG}
\end{equation}
Now tracing over the left wedge kills the off-diagonal terms $k'=k$, and leaves a thermal state in the mode $\hat{b}=\hat{b}^{\mathrm{in}}_{\mathcal{U}_{R}}$:
\begin{equation}
\begin{aligned}
\rho_{R}^{\mathrm{in}}=\mathrm{Tr}_{L}\,\rho_{G}
&=\frac{1}{\cosh^{2}r_{U}}\sum_{k\ge 0}\tanh^{2k}r_{U}\,
|k\rangle\langle k|_{\hat{b}}
\;\otimes\;|0\rangle\langle 0|_{\mathcal{V}^{\mathrm{in}}_{R}}\\[3pt]
&=\sum_{k\ge 0}\frac{N_{U}^{k}}{\left(1+N_{U}\right)^{k+1}}\,
|k\rangle\langle k|_{\hat{b}}
\;\otimes\;|0\rangle\langle 0|_{\mathcal{V}^{\mathrm{in}}_{R}},
\end{aligned}
\label{eq:rhoR}
\end{equation}
with mean occupation $N_{U}=\sinh^{2}r_{U}$. Writing
$\tanh^{2}r_{U}=e^{-2\pi\Omega_{\mathrm{eff}}/a}$ defines an effective Rindler frequency
$\Omega_{\mathrm{eff}}$ which coincides with $|\omega|$ only when $\Delta_{n}\to\infty$ and the
reduced state is thermal in form at all field strengths, but at the Unruh temperature
$a/2\pi$ only in the absence of the electric field.

In the local out-basis the same density matrix reads
\begin{equation}
\rho_{R}^{\mathrm{out}}
=\hat{S}_{AB}(r_{S})
\left[|0\rangle\langle 0|_{A}\otimes\rho_{\mathrm{th}}(N_{U})_{B}\right]
\hat{S}^{\dagger}_{AB}(r_{S}),
\label{eq:rhoRout}
\end{equation}
which is the Schwinger squeezer applied to Eq. \eqref{eq:rhoR}. Since $\hat{S}_{AB}$ is unitary
and acts only on $R$, the spectrum of $\rho_{R}$ is unchanged which is the precise
statement of why the Schwinger effect cannot alter the $R|L$ entanglement.

\paragraph{Entanglement entropy:}
Since Eq. \eqref{eq:rhoR} is thermal, the Von Neumann entropy is the standard bosonic
expression
\begin{equation}
\begin{aligned}
S(R|L)&=-\mathrm{Tr}\,\rho_{R}\log_{2}\rho_{R}
=\left(1+N_{U}\right)\log_{2}\left(1+N_{U}\right)-N_{U}\log_{2}N_{U}\\[3pt]
&=\cosh^{2}r_{U}\log_{2}\cosh^{2}r_{U}-\sinh^{2}r_{U}\log_{2}\sinh^{2}r_{U},
\end{aligned}
\label{eq:Sent}
\end{equation}
or, explicitly in terms of the field strengths,
\begin{equation}
S(R|L)=\frac{X+s}{X-1}\log_{2}\!\left(\frac{X+s}{X-1}\right)
-\frac{1+s}{X-1}\log_{2}\!\left(\frac{1+s}{X-1}\right),
\qquad s=e^{-\pi\Delta_{n}},\ \ X=e^{2\pi|\omega|/a}.
\label{eq:Sentexplicit}
\end{equation}
For $s\to 0$ this is the Bose-Einstein entropy of a mode of frequency $|\omega|$ at
temperature $a/2\pi$. It vanishes as $a\to 0$ and diverges logarithmically as
$|\omega|/a\to 0$. Like Eq. \eqref{eq:ENtotal}, it increases monotonically with the electric
field.

The whole Rényi entropies follows from
$\mathrm{Tr}\,\rho_{R}^{\alpha}=\left[(1+N_{U})^{\alpha}-N_{U}^{\alpha}\right]^{-1}$,
giving
\begin{equation}
S_{\alpha}(R|L)=-\frac{1}{1-\alpha}
\log_{2}\left[\left(1+N_{U}\right)^{\alpha}-N_{U}^{\alpha}\right].
\label{eq:Renyi}
\end{equation}
Setting $\alpha=\tfrac12$ and using
$\left(\sqrt{1+N_{U}}-\sqrt{N_{U}}\right)\left(\sqrt{1+N_{U}}+\sqrt{N_{U}}\right)=1$ gives
\begin{equation}
S_{1/2}(R|L)=2\log_{2}\left(\sqrt{1+N_{U}}+\sqrt{N_{U}}\right)
=\log_{2}e^{2r_{U}}=E_{\mathcal{N}}(R|L),
\label{eq:RenyiHalf}
\end{equation}
which is known that for a pure state the logarithmic negativity is the Rényi-$1/2$
entropy. The euations \eqref{eq:Sent} and \eqref{eq:ENtotal} are
therefore distinct functions that agree only qualitatively, both are monotone measures for
the pure bipartite system.

\paragraph{Negativity:}
The local in-out transformation Eq. \eqref{eq:localbog} acts entirely within $R$, and its
primed counterpart entirely within $L$, both are local Gaussian unitaries and cannot change
entanglement across the $R|L$ partition. The state is pure across that cut and is a
two-mode squeezed vacuum of parameter $r_{U}$, so
\begin{equation}
E_{\mathcal{N}}(R|L)=\frac{2r_{U}}{\ln 2}
=\log_{2}\frac{1+|Z_{\omega}|}{1-|Z_{\omega}|}
=\log_{2}\left[
\frac{\sqrt{X+s}+\sqrt{1+s}}{\sqrt{X+s}-\sqrt{1+s}}\right],
\label{eq:ENtotal}
\end{equation}
or can be written in explicit form as
\begin{equation}
E_{\mathcal{N}}(R|L)=\log_{2}\left[
\frac{e^{2\pi|\omega|/a}+1+2e^{-\pi\Delta_{n}}
+2\sqrt{\left(1+e^{-\pi\Delta_{n}}\right)
\left(e^{2\pi|\omega|/a}+e^{-\pi\Delta_{n}}\right)}}
{e^{2\pi|\omega|/a}-1}\right].
\label{eq:ENtotalfull}
\end{equation}
For $\Delta_{n}\to\infty$ this becomes $\log_{2}\coth\!\left(\pi|\omega|/2a\right)$, the
standard bosonic Unruh entanglement. From Eq. \eqref{eq:monotone}, Eq. \eqref{eq:ENtotal} is
monotonically increasing in the electric field and monotonically decreasing in the magnetic
field. Within the specified family of global in-vacua, the field-dependent squeezing parameter thus changes the $R$-$L$ entanglement. This variation across vacuum states must be distinguished from the invariance of wedge entanglement under local in-out transformations applied to any fixed state.

\subsection{Gaussian Schmidt modes across the Rindler horizon}
\label{sec:schmidt}
The vanishing negativity of the reduced $(B,C)$ state does not imply separability of the complete $(A,B)|(C,D)$ bipartition. This can be seen directly by identifying the collective local modes that carry the horizon entanglement. The mode-wise Schmidt decomposition for a pure bipartite Gaussian state follows from local symplectic transformations \cite{Botero:2003,Serafini:2005}. The related theory of Gaussian entanglement, mixedness and collective-mode correlations is discussed in Refs.~\cite{Adesso:2004,Adesso:2004purity,Adesso:2004multipartite}. In the negative-frequency sector we write the local in-mode operators as $(u,b)$ on $R$ and $(c,d)$ on $L$. With the phases of the local Bogoliubov coefficients absorbed into mode definitions, Eq.~\eqref{eq:localbog} may be expressed as
\begin{equation}
\begin{aligned}
A&=\cosh r_S\,u-\sinh r_S\,b^\dagger,&
B&=\cosh r_S\,b-\sinh r_S\,u^\dagger,\\
C&=\cosh r_S\,c-\sinh r_S\,d^\dagger,&
D&=\cosh r_S\,d-\sinh r_S\,c^\dagger.
\end{aligned}
\label{eq:schmidt_local}
\end{equation}
Consequently, the collective outgoing-mode combinations
\begin{equation}
\begin{aligned}
 b_R&\equiv \cosh r_S\,B+\sinh r_S\,A^\dagger=b,\\
 c_L&\equiv \cosh r_S\,C+\sinh r_S\,D^\dagger=c
\end{aligned}
\label{eq:schmidt_modes}
\end{equation}
recover the two local in-modes mixed by the global vacuum. Their canonical commutators are unchanged, the complementary combinations $u_R=\cosh r_S A+\sinh r_S B^\dagger$ and $d_L=\cosh r_S D+\sinh r_S C^\dagger$ are in their local vacua. The modes $b_R$ and $c_L$ are wedge-local collective modes, not individual outgoing charged-particle modes.

For clarity, consider the quadrature ordering $(A,B)|(C,D)$, and put $\mathbb{I}=\mathbb{I}_2$ and $Z=\mathrm{diag}(1,-1)$. The inverse local two-mode squeeze is represented in either wedge by the symplectic matrix
\begin{equation}
S_{\rm inv}(r_S)=
\begin{pmatrix}
\cosh r_S\,\mathbb{I}&\sinh r_S\,Z\\
\sinh r_S\,Z&\cosh r_S\,\mathbb{I}
\end{pmatrix}.
\label{eq:schmidt_symp}
\end{equation}
After applying $S_R\oplus S_L=S_{\rm inv}(r_S)\oplus S_{\rm inv}(r_S)$ to the full covariance matrix in a consistent phase convention, and reordering the modes as $(b_R,c_L,u_R,d_L)$, the Gaussian Schmidt form is
\begin{equation}
\sigma_{\rm Schmidt}=
\begin{pmatrix}
\cosh(2r_U)\,\mathbb{I}&\sinh(2r_U)\,Z&0&0\\
\sinh(2r_U)\,Z&\cosh(2r_U)\,\mathbb{I}&0&0\\
0&0&\mathbb{I}&0\\
0&0&0&\mathbb{I}
\end{pmatrix}.
\label{eq:schmidt_cov}
\end{equation}
Equivalently, the global state is locally Gaussian-unitarily equivalent to a single two-mode squeezed vacuum of squeezing $r_U$, tensored with two uncorrelated vacua. The two symplectic eigenvalues of the reduced \emph{right-wedge} covariance matrix therefore equal
\begin{equation}
\nu_{R,1}=\cosh(2r_U),\qquad \nu_{R,2}=1.
\label{eq:schmidt_eigs}
\end{equation}
The collective-pair correlations obey
\begin{equation}
\langle b_R^\dagger b_R\rangle=\langle c_L^\dagger c_L\rangle=\sinh^2r_U,
\qquad |\langle b_Rc_L\rangle|=\sinh r_U\cosh r_U.
\label{eq:schmidt_corr}
\end{equation}
It follows that the Schmidt pair carries the full horizon logarithmic negativity,
\begin{equation}
E_{\mathcal N}(b_R:c_L)=E_{\mathcal N}(R:L)=\frac{2r_U}{\ln2},
\label{eq:schmidt_EN}
\end{equation}
whereas tracing out $A,D$ can yield $E_{\mathcal N}(B:C)=0$ as in Eq.~\eqref{eq:threshold}. Thus the apparently missing pairwise entanglement is encoded in collective wedge-local degrees of freedom, including correlations involving the discarded outgoing modes. This is a change of mode representation, not evidence of a conserved transfer of logarithmic negativity between pairs.

The decomposition in Eq.~\eqref{eq:schmidt_cov} follows from the already-established local Gaussian unitary structure of the vacuum; the useful characterization here is the explicit identification of its entanglement-carrying collective outgoing modes. In particular, the Schmidt construction supplies an explanation of reduced-pair separability, not an independent particle-production mechanism or an additional independent entanglement resource.

\subsection{Antiparticle in $R$ with particle in $L$}\label{subs:EN-B-C}

This is the pair $(B,C)$ with the remaining two modes traced out,
\begin{equation}
\rho_{BC}=\mathrm{Tr}_{A,D}\;|0\rangle_{G}\,{}_{G}\langle 0| ,
\label{eq:rhoBC}
\end{equation}
so that, unlike Sec. \ref{sec:EN-RL}, the state is mixed and the von Neumann entropy
is no longer a measure of entanglement.

\paragraph{The reduced state in standard form:}
On deleting the $A$ and $D$ rows and columns of the covariance matrix in Eq. \eqref{eq:covmatrix}
leaves, with the phases absorbed as in Appendix \ref{app:cov} leads to the following
\begin{equation}
\sigma_{BC}=
\begin{pmatrix}
\left(2n+1\right)\mathbbm{1}_{2} & 2m\,\sigma_{z}\\[3pt]
2m\,\sigma_{z} & \left(2n+1\right)\mathbbm{1}_{2}
\end{pmatrix},
\qquad
n=q_{2}+q_{1}N_{U},
\qquad
m=q_{1}M_{U},
\label{eq:sigmaBC}
\end{equation}
which is already in the standard form of a symmetric two-mode squeezed thermal
state. Explicitly,
\begin{equation}
\rho_{BC}=\hat{S}_{BC}(r_{e})
\left[\rho_{\mathrm{th}}(\bar{n})\otimes\rho_{\mathrm{th}}(\bar{n})\right]
\hat{S}^{\dagger}_{BC}(r_{e}),
\qquad
\rho_{\mathrm{th}}(\bar{n})=\sum_{k\ge0}\frac{\bar{n}^{k}}{(1+\bar{n})^{k+1}}
|k\rangle\langle k| ,
\label{eq:rhoBCstandard}
\end{equation}
with the effective squeezing $r_{e}$ and the thermal occupation
$\bar{n}=(\mu-1)/2$ fixed by
\begin{equation}
2n+1=\mu\cosh 2r_{e},
\qquad
2m=\mu\sinh 2r_{e},
\qquad
\mu=\sqrt{\left(2n+1\right)^{2}-4m^{2}},
\qquad
\tanh 2r_{e}=\frac{2m}{2n+1}.
\label{eq:effparams}
\end{equation}

Here $\mu$ is the common symplectic eigenvalue
of $\sigma_{BC}$ and characterizes the mixedness
of the reduced state obtained by tracing out
the modes $A$ and $D$. Inserting Eq. \eqref{eq:moments} gives
\begin{equation}
\mu^{2}=q_{1}^{2}+q_{2}^{2}+2q_{1}q_{2}\cosh 2r_{U}
=q_{1}^{2}+q_{2}^{2}+2q_{1}q_{2}\,\frac{X+1+2s}{X-1},
\qquad
\tanh 2r_{e}=\frac{q_{1}\sinh 2r_{U}}{q_{2}+q_{1}\cosh 2r_{U}} .
\label{eq:muexplicit}
\end{equation}
As $q_{2}\to 0$ one has $\mu\to 1$, $\bar{n}\to 0$ and $r_{e}\to r_{U}$, the reduced state
becomes the pure two-mode squeezed vacuum of Sec. \ref{sec:EN-RL}.

\paragraph{Partial transposition:}
In phase space, transposition of the $C$ subsystem is time reversal on that mode,
$p_{C}\to-p_{C}$, implemented by
$\Lambda=\mathbbm{1}_{2}\oplus\sigma_{z}=\mathrm{diag}(1,1,1,-1)$. Since
$\sigma_{z}\sigma_{z}=\mathbbm{1}_{2}$, the off-diagonal block of Eq. \eqref{eq:sigmaBC} loses
its relative minus sign as
\begin{equation}
\sigma_{BC}\;\longrightarrow\;
\tilde{\sigma}_{BC}=\Lambda\,\sigma_{BC}\,\Lambda=
\begin{pmatrix}
\left(2n+1\right)\mathbbm{1}_{2} & 2m\,\mathbbm{1}_{2}\\[3pt]
2m\,\mathbbm{1}_{2} & \left(2n+1\right)\mathbbm{1}_{2}
\end{pmatrix}.
\label{eq:PTsigma}
\end{equation}
A matrix of the form $\left(\begin{smallmatrix}\lambda\mathbbm{1}_{2}&\eta\mathbbm{1}_{2}\\
\eta\mathbbm{1}_{2}&\lambda\mathbbm{1}_{2}\end{smallmatrix}\right)$ is diagonalised by the
$\pm$ combinations, so the symplectic eigenvalues of Eq. \eqref{eq:PTsigma} are simply
\begin{equation}
\tilde{\nu}_{\pm}=\left(2n+1\right)\pm 2m
=\mu\left(\cosh 2r_{e}\pm\sinh 2r_{e}\right)=\mu\,e^{\pm 2r_{e}} .
\label{eq:PTeigs}
\end{equation}
The state $\tilde{\rho}_{BC}$ is unphysical or equivalently
$\tilde{\sigma}_{BC}+i\Omega\not\ge 0$, precisely when $\tilde{\nu}_{-}<1$, and the logarithmic negativity can be expressed as
\begin{equation}
E_{\mathcal{N}}(B{:}C)=\max\left\{0,\;-\log_{2}\tilde{\nu}_{-}\right\}
=\max\left\{0,\;\frac{2r_{e}}{\ln 2}-\log_{2}\mu\right\}.
\label{eq:ENBCstandard}
\end{equation}
The second expression leads us to infer that entanglement is effective squeezing minus
thermal noise. In the equivalent Fock-space language,
$\tilde{\rho}_{BC}=\left(\mathbbm{1}\otimes T_{C}\right)\rho_{BC}$ has matrix elements
$\langle j_{B}k_{C}|\tilde{\rho}_{BC}|j'_{B}k'_{C}\rangle
=\langle j_{B}k'_{C}|\rho_{BC}|j'_{B}k_{C}\rangle$, and it is the interchange of the
$C$-indices that turns the pair correlator $\langle BC\rangle$ from a phase-coherent
off-diagonal element into a would-be population, forcing a negative eigenvalue whenever
$m>n$.

On substituting Eq. \eqref{eq:moments} into Eq. \eqref{eq:PTeigs} and using the identities
$1-2\sinh r_{U}e^{-r_{U}}=e^{-2r_{U}}$,
\begin{equation}
\tilde{\nu}^{(BC)}_{-}
=e^{-2r_{U}}\cosh^{2}r_{S}+\sinh^{2}r_{S}
=q_{1}\,\frac{1-|Z_{\omega}|}{1+|Z_{\omega}|}+q_{2},
\label{eq:nuBC}
\end{equation}
so that negativity can written as
\begin{equation}
E_{\mathcal{N}}(B{:}C)=\max\left\{0,\;
-\log_{2}\left[\frac{X-1}{X+s}\left(
\frac{X(1+s)}{X+1+2s+2\sqrt{(1+s)(X+s)}}+s\right)\right]\right\}.
\label{eq:ENBC}
\end{equation}

For $r_S=0$, the logarithmic negativity reduces
to $2r_U/\ln 2$, in agreement with Eq. \eqref{eq:ENtotal}, when
$q_{2}=0$ the modes $A$ and $D$ decouple, and the entire right-left wedge entanglement
is carried by the pair $B:C$.
Setting $\tilde{\nu}^{(BC)}_{-}=1$ and using $q_{1}=1+q_{2}$ gives a
remarkably simple separability criterion,
\begin{equation}
(B,C)\ \text{is entangled}
\iff
q_{2}<|Z_{\omega}|
\iff
|\beta_{\omega}|^{2}<\tanh r_{U},
\label{eq:threshold}
\end{equation}
which is given by
\begin{equation}
e^{-2\pi\Delta_{n}}\left(e^{2\pi|\omega|/a}-1\right)^{2}
<\left(1+e^{-\pi\Delta_{n}}\right)
\left(e^{2\pi|\omega|/a}+e^{-\pi\Delta_{n}}\right).
\label{eq:thresholdexplicit}
\end{equation}
The mode pair remains entangled when its local Schwinger occupation $q_2$ remains below the global mixing amplitude $\tanh r_U$. For $|\omega|/a=0.5$ the root of Eq. \eqref{eq:thresholdexplicit} lies at
$s\simeq 0.243$, i.e.\ $\Delta_{n}\simeq 0.45$, for stronger electric fields (or weaker
magnetic fields) the pair becomes positive partial transpose.

\paragraph{Analytical entanglement-separability boundary:}
The two-mode state becomes positive under partial transposition when the smaller symplectic eigenvalue reaches unity. With $s=e^{-\pi\Delta_n}$ and $X=e^{2\pi|\omega|/a}$, the equality $q_2=\tanh r_U$ yields
\begin{equation}
 s^2(X-1)^2=(1+s)(X+s)
 \quad\Longleftrightarrow\quad
 X(X-2)s^2-(X+1)s-X=0.
 \label{eq:phaseboundary_poly}
\end{equation}
For $X>2+\sqrt5$ the unique physical root is
\begin{equation}
 s_c(X)=\frac{X+1+\sqrt{(X+1)^2+4X^2(X-2)}}{2X(X-2)}\,,
 \qquad 0<s_c<1.
 \label{eq:scritical}
\end{equation}
The selected pair is entangled for $0\le s<s_c(X)$ and separable for $s\ge s_c(X)$. For $1<X\le 2+\sqrt5$ there is no separability boundary in the physical interval $0\le s<1$: the pair remains entangled. Equivalently, the threshold first reaches the boundary $s=1$ at
\begin{equation}
 \left(\frac{|\omega|}{a}\right)_{\!c}=\frac{\ln(2+\sqrt5)}{2\pi}\simeq 0.22976.
 \label{eq:nuc}
\end{equation}
This boundary describes separability of a mode-reduced state, rather than a thermodynamic phase transition. Increasing $s$ corresponds, at fixed $M_n$, to increasing $|qE|$. At fixed $|qE|$, increasing $|qB|$ decreases $s$. These statements apply mode by mode, before any integration over the Landau-level degeneracy or frequency spectrum.
The entanglement-separability boundary is displayed in Fig.~\ref{fig:density-three}, together with the other outgoing bipartitions. These results refer to the $\omega<0$ sector.

\paragraph{Distinguishing pairwise from total wedge entanglement:}
The disappearance of $E_{\mathcal{N}}(B{:}C)$ above the boundary does not imply that the four-mode state factorizes across $R|L$. The transformation between local in and out operators acts separately within each wedge, so the full $R|L$ entanglement of the fixed global state is unchanged by these local transformations. In contrast, tracing out $A$ and $D$ discards locally accessible degrees of freedom and can leave the surviving $B:C$ pair separable. This demonstrates the difference between correlations of a selected reduced mode pair and entanglement across the complete wedges. It does not establish a conserved transfer of entanglement between pairs.

\subsection{Particle in $R$ with antiparticle in $L$}\label{sec:AD}
The opposite cross-wedge pairing, $(A,D)$, provides a useful control case. In the global vacuum, its two-mode covariance matrix has equal local occupations $n_A=n_D=q_2(1+N_U)$ and anomalous correlator $|\langle AD\rangle|=q_2 M_U$. Its partially transposed minimum symplectic eigenvalue is therefore
\begin{equation}
\begin{aligned}
\tilde\nu_-^{(AD)}&=1+2q_2(1+N_U-M_U)\\
&=1+q_2(1+e^{-2r_U})\ge1.
\end{aligned}\label{eq:nuAD}
\end{equation}
Thus we have
\begin{equation}
E_{\mathcal{N}}(A{:}D)=0\,,\label{eq:ENAD}
\end{equation}
for the entire parameter domain of this Gaussian construction. For a two-mode Gaussian state, positivity under partial transpose is sufficient for separability, hence the reduced $(A,D)$ pair is separable, not merely undetected by this entanglement measure. The result reflects $|\langle AD\rangle|/n_A=M_U/(1+N_U)=\tanh r_U<1$ whenever $q_2>0$. By contrast, the $(B,C)$ pair has a finite entangled region, Eq.~\eqref{eq:threshold}. The two mode pairings are therefore inequivalent even though both connect the two Rindler wedges.

\subsection{The intra-wedge Schwinger pair}\label{sec:intrawedge}

For the intra-wedge pair $(A,B)$, we must distinguish two states. One is the pure local
in-vacuum on the right wedge. The other is the reduced state obtained from the pure
four-mode global vacuum by tracing out the left wedge. These are different state
preparations and should not be identified with each other.

\paragraph{Which state?}
Relative to the \emph{local} in-vacuum $|0\rangle^{\mathrm{in}}_{R}$ of
Eq. \eqref{eq:schwingersq}, the pair is in the pure two-mode squeezed vacuum generated by the
Schwinger process alone, this is the state relevant to an observer who prepares the Rindler
in-vacuum and asks what the electric field produces in it. Starting from the \emph{pure global vacuum}, by contrast,
\begin{equation}
\rho_{AB}=\mathrm{Tr}_{C,D}\,|0\rangle_{G}\,{}_{G}\langle 0|
=\rho_{R}
\label{eq:rhoAB}
\end{equation}
is the reduced right-wedge state Eq. \eqref{eq:rhoRout}. This reduced $A{:}B$ state is
generally mixed because the pure global vacuum is entangled across the two wedges.
The global vacuum itself remains pure. Its entropy is therefore not an $A|B$
entanglement measure at all, it is the $R|L$ entanglement entropy already computed in
Eq. \eqref{eq:Sent}. We treat the two cases in turn.

\paragraph{(i) The pure Schwinger pair in the local in-vacuum:}
From Eq. \eqref{eq:schwingersq}, the local in-vacuum is
\begin{equation}
|0\rangle^{\mathrm{in}}_{R}
=\frac{1}{\cosh r_{S}}\sum_{l\ge 0}
\left(e^{i\varphi}\tanh r_{S}\right)^{l}\,
|l\rangle_{A}\,|l\rangle_{B},
\qquad
\tanh r_{S}=|Z_{S}|=\sqrt{\frac{q_{2}}{q_{1}}},
\label{eq:schwingerTMSV}
\end{equation}
a pure state across the $A|B$ cut. Tracing over the antiparticle mode gives a thermal
particle distribution,
\begin{equation}
\rho_{A}=\mathrm{Tr}_{B}\,|0\rangle^{\mathrm{in}}_{R}\,{}^{\mathrm{in}}_{R}\langle 0|
=\sum_{l\ge 0}\frac{q_{2}^{\,l}}{q_{1}^{\,l+1}}\,|l\rangle\langle l|_{A}
=\sum_{l\ge 0}\frac{|\beta_{\omega}|^{2l}}{\left(1+|\beta_{\omega}|^{2}\right)^{l+1}}\,
|l\rangle\langle l|_{A},
\label{eq:rhoA}
\end{equation}
with mean occupation $\sinh^{2}r_{S}=q_{2}=|\beta_{\omega}|^{2}$. The entanglement entropy
of the Schwinger pair is therefore
\begin{equation}
\;
\begin{aligned}
S_{S}(A|B)&=q_{1}\log_{2}q_{1}-q_{2}\log_{2}q_{2}\\[3pt]
&=\left(1+|\beta_{\omega}|^{2}\right)\log_{2}\left(1+|\beta_{\omega}|^{2}\right)
-|\beta_{\omega}|^{2}\log_{2}|\beta_{\omega}|^{2},
\end{aligned}\;
\label{eq:SS}
\end{equation}
or it can be written explicitly in terms of the field strengths as
\begin{equation}
S_{S}(A|B)=\frac{X(1+s)}{X+s}\log_{2}\!\left(\frac{X(1+s)}{X+s}\right)
-\frac{s(X-1)}{X+s}\log_{2}\!\left(\frac{s(X-1)}{X+s}\right).
\label{eq:SSexplicit}
\end{equation}
The corresponding Rényi family and negativity follow as in
Eqs. \eqref{eq:Renyi}-\eqref{eq:RenyiHalf} with $N_{U}\to q_{2}$ as
\begin{equation}
S_{\alpha}(A|B)=-\frac{1}{1-\alpha}\log_{2}\left[q_{1}^{\alpha}-q_{2}^{\alpha}\right],
\qquad
E_{\mathcal{N}}(A|B)\Big|_{\mathrm{in}}=\frac{2r_{S}}{\ln 2}
=\log_{2}\left(\sqrt{q_{1}}+\sqrt{q_{2}}\right)^{2}=S_{1/2}(A|B).
\label{eq:SSrenyi}
\end{equation}
Here two limits are worth recording. As $a\to 0$ (that is $X\to\infty$) one has
$q_{2}\to s=e^{-\pi\Delta_{n}}$ and
\begin{equation}
S_{S}(A|B)\;\longrightarrow\;
\left(1+e^{-\pi\Delta_{n}}\right)\log_{2}\left(1+e^{-\pi\Delta_{n}}\right)
-e^{-\pi\Delta_{n}}\log_{2}e^{-\pi\Delta_{n}}
\;\simeq\;\frac{e^{-\pi\Delta_n}}{\ln 2}\left(1+\pi\Delta_n\right),
\label{eq:SSflat}
\end{equation}
the entanglement entropy of a Schwinger pair created in Minkowski space by parallel
electric and magnetic fields. It decreases monotonically with $\Delta_{n}$, so the magnetic
field degrades the pair entanglement through the Landau gap, in agreement with the
inertial-observer analysis of \cite{Bhattacharya:2020sjr}. As $s\to 0$ the entropy vanishes: with no
electric field there is no pair, and $|0\rangle^{\mathrm{in}}_{R}$ is the unentangled
vacuum.

\paragraph{(ii) The reduced Schwinger pair from the pure global vacuum:}
Here $\rho_{AB}=\mathrm{Tr}_{CD}\rho_G$ is mixed and the general formula in Eq. \eqref{eq:nugeneral} of
Appendix~\ref{app:cov} is required, with the moments
\begin{equation}
\begin{aligned}
n_{A}&=\sinh^{2}r_{S}\cosh^{2}r_{U},
&\qquad
n_{B}&=\sinh^{2}r_{S}+\cosh^{2}r_{S}\sinh^{2}r_{U},\\
|\langle AB\rangle|&=\sinh r_{S}\cosh r_{S}\cosh^{2}r_{U} . &&
\end{aligned}
\label{eq:ABmoments}
\end{equation}
A short computation gives the remarkably simple determinant
\begin{equation}
\;
\left(2n_{A}+1\right)\left(2n_{B}+1\right)-4\left|\langle AB\rangle\right|^{2}
=\cosh 2r_{U}=1+2N_{U},
\qquad
\sqrt{\det\sigma_{AB}}=1+2N_{U},\;
\label{eq:detAB}
\end{equation}
so that the symplectic spectrum of $\sigma_{AB}$ is
\begin{equation}
\left\{\nu_{-},\nu_{+}\right\}=\left\{1,\;1+2N_{U}\right\}:
\label{eq:sympAB}
\end{equation}
one exactly pure normal mode and one thermal normal mode of occupation $N_{U}$. This is
just the covariance-matrix restatement of Eq. \eqref{eq:rhoRout}, and it makes two points at
once. First, the mixedness of the reduced $A{:}B$ state obtained from the pure global vacuum,
$\mathrm{Tr}\,\rho_{AB}^{2}=1/(1+2N_{U})$, depends on the global squeezing parameter $r_U$, and is invariant under the local in-out
squeezing for a fixed global input state. Since $r_U$ itself can depend on the electromagnetic
background in this vacuum prescription, this does not imply electric-field independence. Second, the entropy of $\rho_{AB}$ is
\begin{equation}
S\!\left(\rho_{AB}\right)=\left(1+N_{U}\right)\log_{2}\left(1+N_{U}\right)
-N_{U}\log_{2}N_{U}=S(R|L),
\label{eq:SAB}
\end{equation}
identical to Eq. \eqref{eq:Sent}, confirming that it measures the correlation with the left
wedge rather than anything internal to the pair.

For the negativity, writing $\mathcal{C}\equiv\cosh 2r_{U}=1+2N_{U}$ and using
$\sinh^{2}2r_{S}=4q_{1}q_{2}$, the two invariants entering Eq. \eqref{eq:nugeneral} reduce to
\begin{equation}
\det\sigma_{AB}=\mathcal{C}^{2},
\qquad
\tilde{\Sigma}_{AB}=1+\mathcal{C}^{2}+4q_{1}q_{2}\left(1+\mathcal{C}\right)^{2},
\label{eq:ABinvariants}
\end{equation}
so that
\begin{equation}
\;
\tilde{\nu}^{(AB)\,2}_{-}
=\frac{1}{2}\left[\tilde{\Sigma}_{AB}
-\sqrt{\tilde{\Sigma}_{AB}^{2}-4\mathcal{C}^{2}}\right],
\qquad
E_{\mathcal{N}}(A{:}B)=\max\left\{0,-\log_{2}\tilde{\nu}^{(AB)}_{-}\right\}.\;
\label{eq:ENAB}
\end{equation}
For $r_{U}\to 0$ one has $\mathcal{C}\to 1$,
$\tilde{\Sigma}_{AB}\to 2+16q_{1}q_{2}=2\cosh 4r_{S}$ and
$\tilde{\nu}^{(AB)}_{-}\to e^{-2r_{S}}$, recovering
$E_{\mathcal{N}}=2r_{S}/\ln 2=S_{1/2}(A|B)$ of case (i) as it must, since the state becomes
pure. For fixed local squeezing $r_S$, adding thermal occupation to the second local in-mode degrades the pairwise negativity. Along a physical parameter sweep in acceleration, however, both $r_U$ and $r_S$ generally vary; monotonicity therefore must be assessed with the full expressions rather than inferred from $r_U$ alone.

\paragraph{Strict positivity of intra-wedge entanglement:}
For the mixed $A{:}B$ state reduced from the pure global vacuum, the two-mode covariance matrix has a
standard-form correlation block $2|\langle AB\rangle|\sigma_z$ and no
phase-insensitive cross-correlation.  Its Gaussian PPT criterion is
$|\langle AB\rangle|^2>n_A n_B$.  Substituting the moments in
Eq. \eqref{eq:ABmoments} gives the exact identity
\begin{equation}
\;
|\langle AB\rangle|^2-n_A n_B
=q_2(1+N_U)>0\quad (q_2>0).\;
\label{eq:ABstrictPPT}
\end{equation}
Thus, unlike $B{:}C$, the intra-wedge $A{:}B$ pair has no finite
entanglement-separability threshold in this idealized mode-resolved
Gaussian model: its negativity is strictly positive whenever the local
Schwinger squeezing is nonzero.  It tends to zero continuously as
$q_2\to0$. The mixedness induced by tracing out the opposite wedge need
not destroy $A{:}B$ entanglement. This assertion concerns only the
specific global vacuum and Gaussian mode pair considered here.

\paragraph{Physical reading:}

The relative roles of the Schwinger and Unruh mechanisms
depend on the choice of outgoing modes, as discussed in
Secs.~\ref{sec:negativity} and \ref{sec:intrawedge}.
For the cross-wedge pair $(B,C)$, the global two-wedge
squeezing generates entanglement, while the local Schwinger
transformation modifies the correlations between the
selected outgoing modes. As a result, the pair becomes
separable beyond the threshold given by
Eq. \eqref{eq:thresholdexplicit}, even though the complete
right--left wedge state remains entangled.

For the intra-wedge pair $(A,B)$, the situation is
different. The local Schwinger transformation generates
particle--antiparticle entanglement, whereas tracing out
the opposite wedge leaves the pair in a mixed state.
However, as shown by Eq.~\eqref{eq:ABstrictPPT},
the reduced pair remains entangled for any nonzero
$q_2$. Thus, the Schwinger and Unruh mode
transformations affect different outgoing pairs
differently. Although local transformations preserve
the entanglement across the complete $R|L$ partition
for a fixed global state, they can change the
entanglement retained by individual reduced pairs.
This does not imply any conservation of pairwise
entanglement.

\paragraph{Comparison of reduced entanglement measures:}
Figure~\ref{fig:density-three} compares the negativities of two reduced mode-pair states obtained from the same pure four-mode global vacuum, at fixed $(s,|\omega|/a)$. The cross-wedge state is $\rho_{BC}=\mathrm{Tr}_{AD}\rho_G$, whereas the intra-wedge state is $\rho_{AB}=\mathrm{Tr}_{CD}\rho_G$. These reductions retain different correlations, and the resulting negativities need not behave alike. This comparison should not be read as a direct or conserved transfer of logarithmic negativity between the pairs. In particular, the plotted $A{:}B$ negativity is calculated from the \emph{mixed reduced global-vacuum state}, not from the independently prepared pure local in-vacuum.

\paragraph{Local versus global state comparison:}
The pure local-in-vacuum $A{:}B$ state and the reduced $A{:}B$ state derived from the pure global vacuum belong to different preparations. The former is the locally prepared pure two-mode squeezed state, the latter is generally mixed because the left wedge has been traced out. The figures in the main text use the reduced global-vacuum states throughout. Whenever we refer to the pure local in-vacuum, we label it separately and do not treat it as a reduction of $\rho_G$.

\subsection{Quantum mutual information of the reduced outgoing pairs}
\label{sec:mutualinfo}
Logarithmic negativity allows us to identify
entangled outgoing mode pairs and determine
their separability boundaries. However, it
does not characterize all correlations present
in a reduced state. To investigate the total
correlations between the outgoing modes, we
calculate their quantum mutual information,
\begin{equation}
I(X{:}Y)=S(\rho_X)+S(\rho_Y)-S(\rho_{XY}),
\label{eq:mi-definition}
\end{equation}
where $S(\rho)=-\mathrm{Tr}(\rho\log_2\rho)$. The mutual information vanishes for a product state and is non-negative in general. In contrast to the von Neumann entropy of one subsystem of a pure bipartite state, it can be used for the mixed states obtained after tracing out the other outgoing modes. It measures total correlations and should not be regarded as an entanglement measure for these reductions \cite{Henderson:2001,Groisman:2005}. In quantum field theory, mutual information has also been used to compare correlations between separated subsystems without identifying it with a pure-state entanglement entropy \cite{Casini:2007}.

The covariance matrices derived above provide the entropies without constructing the reduced density operators in a number-state basis \cite{Serafini:2004,Weedbrook:2012,Adesso:2014}. For a single bosonic Gaussian mode with mean occupation $n$, the entropy is $g(n)$, where
\begin{equation}
g(n)=(n+1)\log_2(n+1)-n\log_2 n,\qquad g(0)=0.
\label{eq:bosonic-entropy-function}
\end{equation}
For the symmetric cross-wedge state $\rho_{BC}=\mathrm{Tr}_{AD}\rho_G$, the local occupations are $n_B=n_C=q_2+q_1N_U$, and the pair correlation is $m_{BC}=q_1M_U$. The two symplectic eigenvalues of its covariance matrix coincide,
\begin{equation}
\mu_{BC}=\sqrt{(2n_B+1)^2-4|m_{BC}|^2}.
\label{eq:miBC-symplectic}
\end{equation}
Thus $S(\rho_B)=S(\rho_C)=g(n_B)$, while the joint entropy is $S(\rho_{BC})=2g[(\mu_{BC}-1)/2]$. We obtain
\begin{equation}
I(B{:}C)=2g(n_B)-2g\!\left(\frac{\mu_{BC}-1}{2}\right).
\label{eq:miBC}
\end{equation}
The dependence on $m_{BC}$ is important: mutual information is sensitive to correlations that need not be strong enough to violate the positive-partial-transpose criterion.

For the intra-wedge state $\rho_{AB}=\mathrm{Tr}_{CD}\rho_G$, the two symplectic eigenvalues are $1$ and $1+2N_U$. Hence $S(\rho_{AB})=g(N_U)$, whereas the one-mode entropies are $g(n_A)$ and $g(n_B)$, giving
\begin{equation}
I(A{:}B)=g(n_A)+g(n_B)-g(N_U),\qquad
n_A=q_2(1+N_U)=e^{-\pi\Delta_n}.
\label{eq:miAB}
\end{equation}
The subtraction of $g(N_U)$ accounts for the mixedness of $\rho_{AB}$ after the left wedge has been traced out. In particular, $S(\rho_{AB})$ is not an entanglement measure between $A$ and $B$. Both Eqs.~\eqref{eq:miBC} and \eqref{eq:miAB} refer to reductions of the \emph{same pure global vacuum}, rather than to the independently prepared pure local in-vacuum.

The difference between entanglement and total correlations is clearest for the cross-wedge pair. At $|\omega|/a=0.5$, the analytical separability threshold lies at $s_c=e^{-\pi\Delta_n}\simeq0.24356$. At this point,
\begin{equation}
E_{\mathcal N}(B{:}C)=0,\qquad I(B{:}C)\simeq0.26045\ \text{bits}.
\label{eq:mi-threshold-example}
\end{equation}
The pair is therefore separable but not in a product state. Increasing $s$ beyond the threshold removes its logarithmic negativity, while the correlations contributing to $I(B{:}C)$ need not disappear. We emphasize that nonzero mutual information
does not determine whether the remaining
correlations are classical, quantum, or a
combination of both. A separate calculation
of quantum discord would be required to
distinguish these contributions
\cite{Ollivier:2001,Modi:2012}.

Two limits provide checks on these expressions and help separate the two physical effects. In the weak-Schwinger limit $s\to0$ at fixed $|\omega|/a>0$, we have $q_2\to0$ and $N_U\to (X-1)^{-1}$. The $B{:}C$ pair approaches a pure two-mode squeezed state, while the $A{:}B$ state approaches the two-mode vacuum. Consequently,
\begin{equation}
I(B{:}C)\longrightarrow 2g\!\left(\frac{1}{X-1}\right),
\qquad I(A{:}B)\longrightarrow0.
\label{eq:mi-weak-schwinger}
\end{equation}
In the opposite large-frequency limit $|\omega|/a\to\infty$ at fixed $s$, the cross-wedge squeezing vanishes, $N_U\to0$, whereas $q_2\to s$. We then find
\begin{equation}
I(B{:}C)\longrightarrow0,\qquad
I(A{:}B)\longrightarrow2g(s).
\label{eq:mi-high-frequency}
\end{equation}
The intra-wedge result is consistent with a pure Schwinger-squeezed pair of mean occupation $s$ in this limiting prescription. These limiting relations concern the specified mode family: they do not, by themselves, establish an operational inertial detector limit.

Figure~\ref{fig:neg-mi} compares mutual information with logarithmic negativity for both reduced pairs. Its purpose is not to assign an entanglement interpretation to mutual information, but to show explicitly that the disappearance of pairwise entanglement and the disappearance of total correlations are different statements.

\subsection{Numerical results and field dependence}
\label{sec:numerical-results}
We now plot the analytical expressions in terms of $e^{-\pi\Delta_n}$ and $|\omega|/a$. The first parameter combines the mass and transverse Landau gap with the electric field, while the second sets the dimensionless Rindler frequency. All plots in this section refer to the negative-frequency sector and to the same specified global vacuum.

\begin{figure*}[t]
\centering
\includegraphics[width=0.99\textwidth]{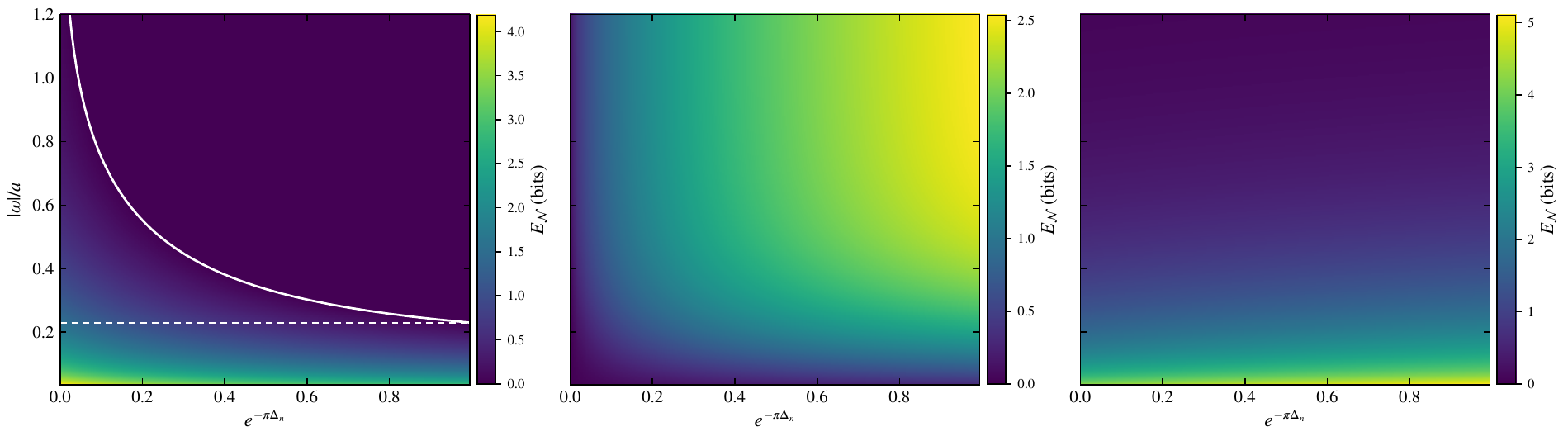}
\caption{\it \small
Logarithmic negativity as a function of
$e^{-\pi\Delta_n}$ and $|\omega|/a$.
From left to right, the panels show the cross-wedge
pair $B{:}C$, the intra-wedge pair $A{:}B$, and
the complete wedge partition $R{:}L$.
Each panel has its own color scale because the
three negativities have different ranges.
In the left panel, the solid white curve marks
the separability boundary $q_2=\tanh r_U$, while
the dashed line indicates the critical value
$|\omega|/a=\ln(2+\sqrt{5})/(2\pi)$.
All panels correspond to the same pure four-mode
global vacuum. The $A{:}B$ and $B{:}C$ panels
represent different mixed reduced states,
obtained by tracing out the complementary modes
$C,D$ and $A,D$, respectively.
Neither panel represents an independently
prepared pure local in-vacuum.
The color scales are restricted to finite ranges
for visualization, as the cross-wedge and
complete-wedge negativities diverge in the
limit $|\omega|/a\to0^+$.
}
\label{fig:density-three}
\end{figure*}

Figure~\ref{fig:density-three} shows three different views of the same state. A finite separability domain appears for the outgoing $B{:}C$ pair, even though the intra-wedge pair has nonzero negativity away from the zero-Schwinger limit and the complete right-left partition remains entangled. Since the full state is pure, none of these observations implies a loss of the entire horizon entanglement.

\begin{figure*}[t]
\centering
\includegraphics[width=0.91\textwidth]{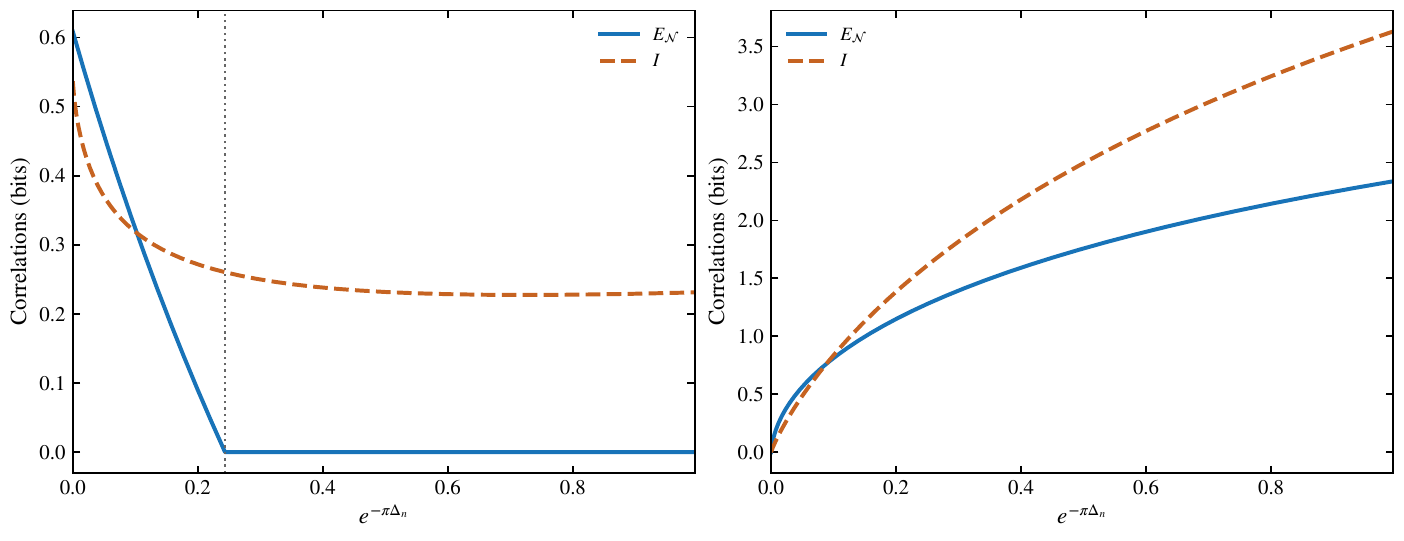}
\caption{ \it \small Logarithmic negativity (solid curves) and quantum mutual information (dashed curves) as functions of $e^{-\pi\Delta_n}$ at fixed $|\omega|/a=0.5$. The left panel concerns the cross-wedge $B{:}C$ pair, and the right panel the intra-wedge $A{:}B$ pair. The dotted vertical line on the left marks the analytical separability threshold, $e^{-\pi\Delta_n}\simeq0.2436$. Mutual information remains nonzero even when the $B{:}C$ negativity vanishes. All quantities are in bits. Both panels use mixed reduced mode-pair states of the same pure four-mode global vacuum, rather than the pure local in-vacuum.}
\label{fig:neg-mi}
\end{figure*}

The line plots in Fig.~\ref{fig:neg-mi} make the correlation distinction particularly clear. Beyond the $B{:}C$ separability threshold, the pair is no longer entangled but is not statistically independent. By contrast, the local $A{:}B$ pair is entangled for nonzero Schwinger squeezing. These statements concern different partitions and do not represent a conserved transfer of logarithmic negativity between pairs.

\begin{figure}[t]
\centering
\includegraphics[width=0.70\linewidth]{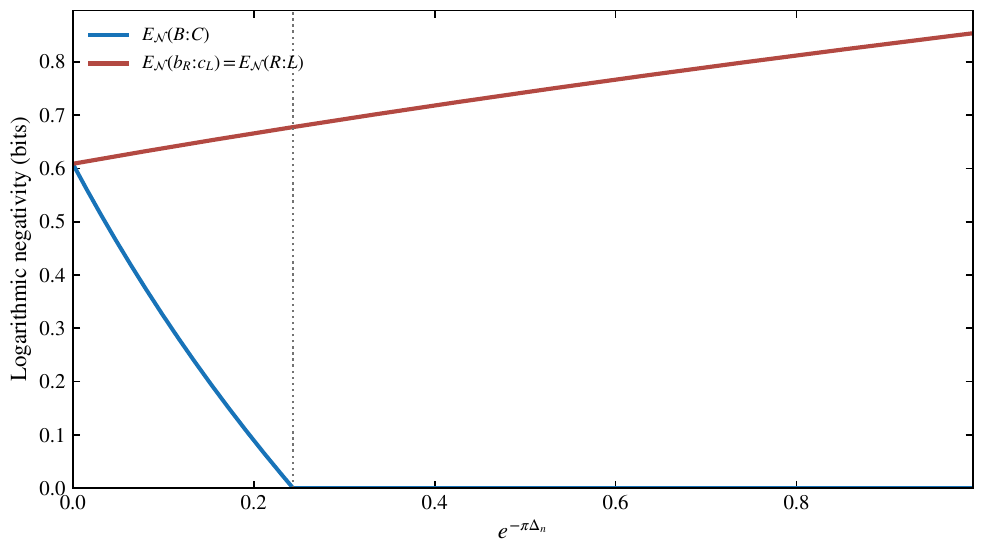}
\caption{ \it \small Entanglement of the selected physical $B{:}C$ pair compared with that of the collective Gaussian Schmidt pair $b_R{:}c_L$ at fixed $|\omega|/a=0.5$. The Schmidt-pair negativity equals the complete $R{:}L$ negativity of the pure global vacuum, $2r_U/\ln2$, whereas the mixed reduced outgoing $B{:}C$ pair becomes separable at the vertical dotted line. The equality of the collective and full-wedge quantities follows from the local Gaussian Schmidt construction, not from a separate entanglement resource.}
\label{fig:schmidt-new}
\end{figure}

Figure~\ref{fig:schmidt-new} gives the mode interpretation of the result. The local inverse Schwinger transformations reconstruct one collective mode in each wedge. Those collective modes retain all the entanglement of the complete wedge partition, although tracing out the remaining outgoing modes can leave $B{:}C$ separable.

\begin{figure*}[t]
\centering
\includegraphics[width=0.97\textwidth]{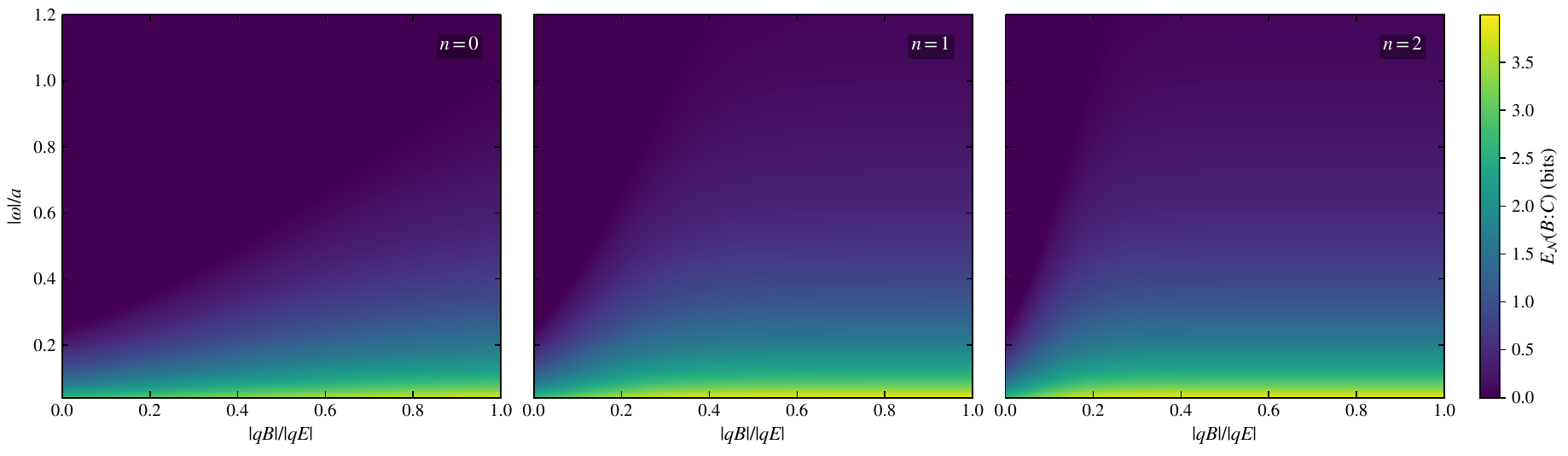}

\caption{\it \small
Cross-wedge logarithmic negativity
$E_{\mathcal N}(B{:}C)$ as a function of
$|qB|/|qE|$ and $|\omega|/a$.
The panels correspond to Landau-level indices
$n=0,1,2$, at fixed $m^2/|qE|=0.05$.
A common color scale is used because all
three panels represent the same entanglement
measure.
Higher Landau levels increase
$\Delta_n=[m^2+(2n+1)|qB|]/|qE|$ and
suppress the Schwinger factor
$s=e^{-\pi\Delta_n}$, thereby shifting the
separability boundary.
At fixed electric field and Rindler frequency,
increasing the magnetic field can restore
the entanglement of a previously separable
$B{:}C$ pair.
The $|qB|/|qE|\to0$ endpoint represents
the formal limit at fixed Landau-level index;
the complete zero-magnetic-field theory
requires a continuous transverse-momentum
description.
}
\label{fig:landau-density}
\end{figure*}

Finally, Fig.~\ref{fig:landau-density} displays the magnetic-field dependence explicitly. The transverse field enters through the Landau gap. Increasing $|qB|$ at fixed $|qE|$ therefore changes the selected pair's negativity by reducing the Schwinger factor: it does not, by itself, establish a change of the universal Unruh temperature.

\subsection{Recovery of the free-field limit}
\label{sec:freefield}

As a consistency check we verify that the whole of Sec.~\ref{sec:negativity}
collapses to the standard neutral-scalar Rindler results when the background fields are
switched off. Two distinct limits are involved, $E\to 0$ and $B\to 0$, and it is instructive
to take them in that order.

\paragraph{Entanglement entropy:}
Setting $E\to 0$ at fixed $B$ sends $\Delta_{n}=M_{n}^{2}/qE\to\infty$, hence $s\to 0$ and
\begin{equation}
N_{U}=\frac{1+s}{X-1}\;\longrightarrow\;n_{\omega}\equiv
\frac{1}{e^{2\pi|\omega|/a}-1},
\label{eq:NUfree}
\end{equation}
which is the Bose-Einstein occupation at the Unruh temperature $T=a/2\pi$. Equation
\eqref{eq:Sent} then becomes
\begin{equation}
\begin{aligned}
S(R|L)\;&\longrightarrow\;
\left(1+n_{\omega}\right)\log_{2}\left(1+n_{\omega}\right)
-n_{\omega}\log_{2}n_{\omega}\\[3pt]
&=-\log_{2}\!\left(1-e^{-2\pi\nu}\right)
+\frac{1}{\ln 2}\,\frac{2\pi\nu}{e^{2\pi\nu}-1},
\qquad \nu=\frac{|\omega|}{a},
\end{aligned}
\label{eq:Sfree}
\end{equation}
which is precisely the per-mode entanglement entropy of a free scalar across the Rindler
horizon, as follows from the Bisognano-Wichmann theorem and the thermal form
$\rho_{R}=e^{-2\pi K}/Z$ with $K$ the boost generator.

Note that $N_U$ in Eq.~\eqref{eq:NUfree} is independent
of the Landau-level index $n$. In the absence of
the electric field, the global squeezing parameter
reduces to the standard bosonic Unruh value, and
the entanglement entropy per mode becomes independent
of the magnetic field. However, the magnetic field
still modifies the transverse energy spectrum and
the degeneracy of the Landau levels, given by
$|qB|/(2\pi)$ per unit transverse area.
Thus, while the magnetic field affects the overall
mode spectrum, it does not change the entanglement
entropy of an individual mode in this limit.
As discussed in Sec.~\ref{sec:leveldep},
the Landau-level dependence of the global squeezing
parameter arises from its combined dependence on
the electric and magnetic fields.

\paragraph{Switching off the magnetic field:}

When the magnetic field is switched off,
the discrete Landau-level spectrum approaches
a continuous transverse-momentum spectrum.
This continuum limit is obtained by taking
$B\to0$ while allowing the Landau-level index
$n$ to increase such that
$(2n+1)|qB|$ remains finite. Consequently,
the Landau-level sum, including its degeneracy,
approaches the transverse momentum integral,
\begin{equation}
\frac{|qB|}{2\pi}\sum_{n=0}^{\infty}
\longrightarrow
\int\frac{d^2k_\perp}{(2\pi)^2},
\qquad
M_n^2\longrightarrow m^2+k_\perp^2.
\label{eq:Bzero}
\end{equation}

The order of the two limits is immaterial for the entropy per mode, since $N_{U}$ has
already lost its $n$-dependence at $s=0$. It does matter for the total entropy, which
becomes the usual transverse-area times ultraviolet-divergent Rindler entanglement entropy,
the divergence resides in the mode sum Eq. \eqref{eq:Bzero}, not in Eq. \eqref{eq:Sfree}.

\paragraph{Other measures:}
At $s=0$ every result of this section degenerates coherently:
\begin{itemize}
\item $q_{2}\to 0$ and $q_{1}\to 1$, so $r_{S}\to 0$, the Schwinger squeezer
$\hat{S}_{AB}\to\mathbbm{1}$, and $\rho_{R}^{\mathrm{out}}=\rho_{R}^{\mathrm{in}}$ in
Eq. \eqref{eq:rhoRout}.
\item $\mu\to 1$, $\bar{n}\to 0$ and $r_{e}\to r_{U}$ in Eq. \eqref{eq:muexplicit}, so
$\rho_{BC}$ becomes pure and
$E_{\mathcal{N}}(B{:}C)=E_{\mathcal{N}}(R|L)=\log_{2}\coth\left(\pi\nu/2\right)$.
\item $\rho_{AD}$ becomes the vacuum and $E_{\mathcal{N}}(A{:}D)=0$ trivially, rather than
through the cancellation of Sec.~\ref{sec:AD}.
\item $E_{\mathcal{N}}(A{:}B)\to 0$: the intra-wedge Schwinger pair disappears.
\end{itemize}
The one measure that does not coincide with the negativity in this limit is the
entropy itself, since $E_{\mathcal{N}}=S_{1/2}\neq S_{1}$ by Eq. \eqref{eq:RenyiHalf} even for
the free field.

\subsection{Summary of limits}

\begin{table}[h]
\centering
\begin{tabular}{lccc}
\toprule
Limit & $E_{\mathcal{N}}(R|L)$ & $E_{\mathcal{N}}(B{:}C)$ & $E_{\mathcal{N}}(A{:}D)$\\
\midrule
$B\to\infty$ or $E\to 0$ \ ($s\to 0$)
& $\log_{2}\coth\frac{\pi|\omega|}{2a}$
& $\log_{2}\coth\frac{\pi|\omega|}{2a}$ & $0$\\[3pt]
$a\to 0$ \ ($X\to\infty$) & $0$ & $0$ & $0$\\[3pt]
$qE\gg M_{n}^{2}$ \ ($s\to 1$)
& $\log_{2}\dfrac{\sqrt{X+1}+\sqrt{2}}{\sqrt{X+1}-\sqrt{2}}$
& $0$ beyond threshold & $0$\\
\bottomrule
\end{tabular}
\caption{Limiting behaviour of the logarithmic negativities in the negative-frequency sector $\omega<0$.}
\label{tab:limits}
\end{table}

The limits in Table~\ref{tab:limits} refer to the
negative-frequency sector $\omega<0$.
At fixed $|\omega|>0$ and fixed background field
strengths, $a\to0$ implies $X\to\infty$ and
$r_U\to0$ in this sector. Consequently, the
logarithmic negativities across the $R|L$
partition and between the selected cross-wedge
mode pairs vanish, whereas the out-mode
occupations in Eqs.~\eqref{eq:nU} and
\eqref{eq:nV} approach the Schwinger mean
occupation $s=e^{-\pi\Delta_n}$.

In the strong-electric-field limit $s\to1$,
at fixed $|\omega|/a$, the entanglement of the
cross-wedge pair $B:C$ depends on the Rindler
frequency. In particular, the pair remains
entangled for
$|\omega|/a<\ln(2+\sqrt{5})/(2\pi)$
and becomes separable at or above this
critical value. This further shows that
the separability of the selected outgoing
pair depends on both the Schwinger parameter
and the Rindler frequency.

This conclusion does not extend to the positive-frequency sector.
Equation~\eqref{eq:ENtotalpos} in Appendix~\ref{app:omegapos}, together
with the corresponding mixing angle, gives
\begin{equation}
\tanh^2 r_U^{(+)}
\xrightarrow[a\to0]{}\frac{s}{1+s},
\qquad
E_{\mathcal N}^{(+)}(R|L)
\xrightarrow[a\to0]{}
\frac{2}{\ln2}\operatorname{arsinh}\sqrt{s}.
\label{eq:positive_frequency_a_limit}
\end{equation}
These expressions describe limits of the mode-resolved quantities
at fixed Rindler frequency and fixed $s$. Their interpretation as
an inertial limit requires examining the limiting mode functions
and their relation to fixed physical wave packets. The vanishing
cross-wedge entanglement found for $\omega<0$ should therefore not
be stated as a result for both frequency sectors.

Within the negative-frequency sector, increasing the electric field
at fixed mass, magnetic field, Landau level, and $|\omega|/a$ increases
$s$ and the total $R|L$ logarithmic negativity through the
field-dependent global mixing angle. In contrast, the reduced
$(B,C)$ pair can become separable when the condition in
Eq.~\eqref{eq:threshold} is no longer satisfied. Thus the total
wedge entanglement and the entanglement of a selected mode pair
can respond differently to the background fields. The local
in-out transformations preserve the total $R|L$ entanglement of
a fixed state, discarding modes can reduce the entanglement
available in a selected pair.

\section{Discussion: Mode Correlations and Physical Implications}
\label{sec:discussion}

Our results show that the entanglement between individual outgoing
particle-antiparticle modes can behave differently from the
entanglement between the complete right and left Rindler wedges.
In particular, the logarithmic negativity of the cross-wedge pair
$B:C$ can vanish beyond a certain value of the Schwinger parameter,
while its mutual information remains nonzero. At the same time,
the complete right-left wedge system remains entangled.
Therefore, the disappearance of entanglement between a particular
pair of outgoing modes does not imply that all correlations across
the Rindler horizon have disappeared.

This difference can be understood from the Gaussian Schmidt
decomposition of the pure global vacuum. The local in-out
Bogoliubov transformations act separately within the two Rindler
wedges and therefore do not change the entanglement across the
complete wedge partition for a fixed global state. The Schmidt
decomposition shows that the cross-wedge entanglement is carried
by a pair of collective modes, while the remaining two modes
are in their local vacua in the transformed basis. Consequently,
tracing out some of the outgoing modes can make a particular
reduced pair separable, even though the complete state remains
entangled across the two wedges.

The correlations between modes associated with different regions
of spacetime have also been investigated in analogue-gravity
experiments. For example, correlations between Hawking radiation
and its partner modes have been observed in analogue black-hole
systems~\cite{Steinhauer:2016}. Quantum simulations have also
been used to study Unruh-like radiation~\cite{Hu:2019} and
Schwinger pair production~\cite{Martinez:2016}. Although these
studies involve physical systems different from the charged
scalar field considered here, they illustrate the broader
interest in understanding particle production and the
correlations between the created modes. A direct comparison
with such experiments would require identifying the
corresponding mode transformations and measuring the
relevant output correlations.

It is also useful to distinguish our results from the
entanglement thresholds obtained for Schwinger pair production
with thermal initial states~\cite{Kranas:2026}. In that work,
the entanglement depends on the thermal preparation of the
initial state, whereas in our analysis the reduced states
are obtained by tracing out selected modes of a specified
pure global vacuum. Thus, the two calculations describe
different physical situations, although both demonstrate
that particle production and entanglement need not follow
the same behaviour.

Finally, our analysis is based on prescribed classical
electromagnetic fields and a particular choice of global
vacuum and mode sectors. The field dependence of the
global-mode squeezing parameter $r_U$ describes the
correlations of this state and should not be interpreted
as a modification of the universal Unruh temperature.
Such an interpretation would require an independent
calculation of the response of a uniformly accelerated
detector. Our conclusions are therefore restricted to
the mode correlations and bipartitions considered here.

\section{Conclusions}
\label{sec:conclusions}

In this work, we have investigated particle production and
quantum correlations of a charged complex scalar field in
Rindler spacetime in the presence of parallel background
electric and magnetic fields. The magnetic field introduces
Landau quantization, which modifies the effective mass of
the scalar field as $M_n^2=m^2+(2n+1)|qB|$. 
Using the established charged-scalar Rindler mode
construction, extended to include the Landau-level
dependence, we obtained the corresponding
Bogoliubov relations and investigated the
entanglement properties of the outgoing
particle--antiparticle modes.
In particular, we considered a pure global vacuum in the
negative-frequency sector and calculated the logarithmic
negativity and mutual information of its reduced states.

We found that the entanglement between the outgoing modes
depends strongly on the choice of the particle-antiparticle
pair. For the cross-wedge pair $B{:}C$, the logarithmic
negativity is nonzero only when $q_2<\tanh r_U$, where $q_2$
characterizes the local Schwinger mixing and $r_U$ is the
global squeezing parameter. This condition gives an
analytical entanglement-separability boundary,
$X(X-2)s^2-(X+1)s-X=0$, where
$s=e^{-\pi\Delta_n}$ and $X=e^{2\pi|\omega|/a}$.
The corresponding critical frequency-to-acceleration
ratio is $\ln(2+\sqrt{5})/(2\pi)$. Our numerical results
show how the entanglement region changes with the
Schwinger parameter and the Rindler frequency.
In contrast, the complementary cross-wedge pair $A{:}D$
remains separable, whereas the intra-wedge pair $A{:}B$
is entangled whenever the local Schwinger squeezing is
nonzero. These results show that different outgoing
particle-antiparticle pairs can have different
entanglement properties even when their reduced states
are obtained from the same pure global vacuum.

We also studied the quantum mutual information between
the outgoing modes to understand their total correlations.
For the $B{:}C$ pair, we found that the disappearance of
logarithmic negativity does not necessarily imply the
absence of correlations. For example, at
$|\omega|/a=0.5$, the pair becomes separable at
$s\simeq0.2436$, while its mutual information remains
finite, $I(B{:}C)\simeq0.2604$ bits. This demonstrates
that the conditions for entanglement and total
correlations are different. Since mutual information
includes both classical and quantum contributions, its
nonzero value in a separable state does not by itself
determine the nature of the remaining correlations.

Another important result concerns the entanglement
between the complete right and left Rindler wedges.
Using the Gaussian Schmidt decomposition of the pure
global vacuum, we identified a pair of collective modes
that carries the entanglement across the two wedges.
The remaining two modes are in their local vacuum states
in the transformed basis. We found that the logarithmic
negativity of this collective pair is equal to that of
the complete wedge partition,
$E_{\mathcal N}(R{:}L)=2r_U/\ln 2$. This quantity can
remain nonzero even when the reduced cross-wedge pair
$B{:}C$ becomes separable. Thus, the loss of
entanglement between a particular pair of outgoing
modes does not imply that the complete right-left
wedge state is separable.

Finally, we examined how the background electric and
magnetic fields affect these correlations. The electric
field enters through the Schwinger parameter, while the
magnetic field changes the effective mass through the
Landau-level spectrum. Consequently, the entanglement
regions depend on both fields and on the Landau-level
index. We emphasize that the field dependence of the
global squeezing parameter is associated with the
particular charged-field vacuum considered here and
should not be interpreted as a change in the universal
Unruh temperature. Our calculations are restricted
to prescribed classical fields and the selected
negative-frequency mode sector. A direct connection
with experimental measurements would require a
description in terms of normalized wave packets and
a suitable detector or quantum simulator.

Overall, our analysis shows that the Schwinger and
Unruh mode transformations lead to different
correlation properties for different outgoing
particle-antiparticle pairs. Although the
entanglement of a selected reduced pair can vanish,
total correlations may remain, and the complete
right-left Rindler-wedge state can still be
entangled through collective modes.
It would also be interesting to extend the present
analysis to localized wave packets and investigate
how the correlations obtained here can be accessed
by uniformly accelerated detectors. Such studies
may provide a better understanding of the relation
between particle production and quantum correlations
in the presence of external electromagnetic fields.
\begin{acknowledgments}
S. K. acknowledges VIT, Vellore, for financial support through the Faculty Seed Grant (RGEMS) scheme. This work was financially supported by Vellore Institute of Technology (VIT), Vellore under the Faculty Seed Grant (RGEMS) (Sanction Order No.: SG20260125). H.P. was supported by the National Research Foundation of Korea (NRF) grant funded by the Korean government (MSIT) (RS-2022-NR069038). H.P. was supported by Brain Pool program funded by the Ministry of Science and ICT through the National Research Foundation of Korea (RS-2025-25457100).
\end{acknowledgments}
\appendix

\section{Mode normalisations}
\label{app:norm}

\subsection*{The Klein-Gordon product}
With the convention $D_\mu=\partial_\mu-iqA_\mu$ and metric signature $(-,+,+,+)$, the conserved Klein-Gordon product on a future-oriented constant-$t$ slice of the right Rindler wedge is
\begin{equation}
(\phi,\psi)_{\rm KG}=i\int_{-\infty}^{\infty}dx\,dy\,dz\,
\bigl[\phi^*D_t\psi-(D_t\phi)^*\psi\bigr].
\label{eq:KGexplicit}
\end{equation}
Indeed, $\sqrt{h}=e^{ax}$ and $n^t=e^{-ax}$, so $\sqrt{h}\,n^t=1$. For $\phi=\psi\propto e^{-i\omega t}$, the integrand equals $2(\omega+qA_t)|\phi|^2$. The sign of the integrated norm is fixed by the corresponding mode's boundary conditions and fluxes, not solely by the sign of $\omega$. On the left wedge the future-directed Killing vector has the opposite orientation. The transverse normalization separates from the longitudinal Klein-Gordon flux normalization, which follows the charged-scalar Rindler mode analysis of \cite{Gabriel:1999yz}.

\subsection*{Transverse (Landau) factor} The equation \eqref{eq:landau} is a harmonic-oscillator problem centred at $y_{0}=k_{z}/(qB)$,
with well-behaved solutions only for $\lambda_{n}=|qB|(2n+1)$. The normalised eigenfunctions
are given by Eq. \eqref{eq:landaufn}, satisfying
$\int dy\,Y_{n}^{*}(y)Y_{n'}(y)=\delta_{nn'}$.

\subsection*{Longitudinal factor}

Using the asymptotic behaviour of the Whittaker functions,
\begin{equation}
W_{\kappa,\mu}(z)\underset{z\to\infty}{\simeq}e^{-z/2}z^{\kappa},
\qquad
W_{\kappa,\mu}(z)\underset{z\to 0}{\simeq}
\frac{\Gamma(2\mu)}{\Gamma\!\left(\tfrac{1}{2}+\mu-\kappa\right)}z^{1/2-\mu}
+\frac{\Gamma(-2\mu)}{\Gamma\!\left(\tfrac{1}{2}-\mu-\kappa\right)}z^{1/2+\mu},
\label{eq:whitasymp}
\end{equation}
and evaluating Eq. \eqref{eq:KGexplicit} on a constant-$t$ slice, one obtains
\begin{align}
\left\langle\mathcal{U}^{\omega}_{\mathrm{in},R},
\mathcal{U}^{\omega'}_{\mathrm{in},R}\right\rangle
&=\left|\mathcal{N}(\mathcal{U}^{\omega}_{\mathrm{in},R})\right|^{2}
\left(\frac{qE}{a}\right)
\frac{e^{\pi\omega/2a}}{\sinh(\pi\omega/a)}\nonumber\\
&\qquad\qquad\times
\cosh\!\left[\pi\left(\frac{\omega}{a}-\frac{\Delta_{n}}{2}\right)\right]
\delta(\omega-\omega')\,\delta(k_{z}-k'_{z})\,\delta_{nn'},
\label{eq:normU}\\[4pt]
\left\langle\mathcal{V}^{\omega}_{\mathrm{in},R},
\mathcal{V}^{\omega'}_{\mathrm{in},R}\right\rangle
&=\left|\mathcal{N}(\mathcal{V}^{\omega}_{\mathrm{in},R})\right|^{2}
\left(\frac{\omega/a}{\sinh(\pi\omega/a)}\right)\left(\frac{qE}{a}\right)
e^{\frac{\pi}{2}\left(\frac{\omega}{a}+\Delta_{n}\right)}\nonumber\\
&\qquad\qquad\times
\cosh\!\left[\pi\left(\frac{\omega}{a}-\frac{\Delta_{n}}{2}\right)\right]
\delta(\omega-\omega')\,\delta(k_{z}-k'_{z})\,\delta_{nn'},
\label{eq:normV}
\end{align}
which are fixed by demanding
\begin{equation}
Q\!\left(\mathcal{U}^{\omega}_{\mathrm{in},R}\right)=\mathrm{sgn}\,\omega=\pm 1,
\qquad
Q\!\left(\mathcal{V}^{\omega}_{\mathrm{in},R}\right)=+1 .
\label{eq:normcond}
\end{equation}
In addition,
\begin{equation}
\left|\mathcal{N}(\mathcal{V}^{\omega}_{\mathrm{out},R})\right|
=\left|\mathcal{N}(\mathcal{U}^{\omega}_{\mathrm{in},R})\right|,
\qquad
\left|\mathcal{N}(\mathcal{U}^{\omega}_{\mathrm{out},P})\right|^{-2}
=|\omega|\left(\frac{qE}{2a^{2}}\right)e^{-\pi\omega/2a},
\label{eq:normmisc}
\end{equation}
the second holding on the past quadrant.

\section{The four-mode covariance matrix}
\label{app:cov}

\subsection{Second moments}

In $|0\rangle_G$, the nonzero independent
second moments, together with the vacuum
occupations, are
\begin{equation}
\begin{aligned}
\langle \hat u^\dagger \hat u\rangle
&=0,
&
\langle \hat w'^\dagger \hat w'\rangle
&=0,
\\
\langle \hat u'^\dagger \hat u'\rangle
&=N_U,
&
\langle \hat w^\dagger \hat w\rangle
&=N_U,\quad
\langle \hat u'\hat w\rangle
=e^{i\theta_\omega}M_U.
\end{aligned}
\label{eq:in_second_moments}
\end{equation}
The Hermitian-conjugate correlator follows
immediately, while the remaining independent
normally ordered second moments vanish.
The reverse-ordered moments are determined by
the bosonic commutation relations. On inserting the inverse Bogoliubov relations in
Eq. \eqref{eq:localbogInv} and their $L$-counterparts, and using the phase constraint
Eq. \eqref{eq:localbogcoef} to align phases, one obtains
\begin{equation}
\begin{aligned}
n_{A}&=q_{2}\left(1+N_{U}\right), &\qquad
n_{B}&=q_{2}+q_{1}N_{U}, \\
n_{C}&=q_{2}+q_{1}N_{U}, &\qquad
n_{D}&=q_{2}\left(1+N_{U}\right),
\end{aligned}
\label{eq:appn}
\end{equation}
and the nonvanishing correlators
\begin{equation}
\begin{aligned}
|\langle AB\rangle|&=|\langle CD\rangle|=\sqrt{q_{1}q_{2}}\left(1+N_{U}\right)
&&\text{(intra-wedge Schwinger pairs)},\\
|\langle AD\rangle|&=q_{2}M_{U},
\qquad
|\langle BC\rangle|=q_{1}M_{U}
&&\text{(cross-wedge pairs)},\\
|\langle A^{\dagger}C\rangle|&=|\langle B^{\dagger}D\rangle|=\sqrt{q_{1}q_{2}}\,M_{U}
&&\text{(cross-wedge number-type)},\\
\langle AC\rangle&=\langle BD\rangle=\langle A^{\dagger}D\rangle
=\langle B^{\dagger}C\rangle=0 .
\end{aligned}
\label{eq:appcorr}
\end{equation}
Note that $|\langle AD\rangle|/n_{A}=M_{U}/(1+N_{U})=\tanh r_{U}=|Z_{\omega}|<1$, which is
the origin of Eq. \eqref{eq:ENAD}.

\subsection{Quadrature form}

Define the quadratures and the phase-space vector as
\begin{equation}
x_{k}=\frac{X_{k}+X_{k}^{\dagger}}{\sqrt{2}},
\qquad
p_{k}=\frac{X_{k}-X_{k}^{\dagger}}{i\sqrt{2}},
\qquad
\xi=\left(x_{A},p_{A},x_{B},p_{B},x_{C},p_{C},x_{D},p_{D}\right)^{T},
\label{eq:quadratures}
\end{equation}
with $\sigma_{jk}=\left\langle\{\Delta\xi_{j},\Delta\xi_{k}\}\right\rangle$, normalised so
that $\sigma_{\mathrm{vac}}=\mathbbm{1}_{8}$. With all phases absorbed into the mode definitions
so that the correlators in Eq. \eqref{eq:appcorr} are real and positive, the covariance matrix has
the following block form
\begin{equation}
\setlength{\arraycolsep}{3pt}
\sigma=\begin{pmatrix}
\left(2n_{A}+1\right)\mathbbm{1}_{2} & 2m_{AB}\,\sigma_{z}
& 2\kappa_{AC}\,\mathbbm{1}_{2} & 2m_{AD}\,\sigma_{z}\\[4pt]
2m_{AB}\,\sigma_{z} & \left(2n_{B}+1\right)\mathbbm{1}_{2}
& 2m_{BC}\,\sigma_{z} & 2\kappa_{BD}\,\mathbbm{1}_{2}\\[4pt]
2\kappa_{AC}\,\mathbbm{1}_{2} & 2m_{BC}\,\sigma_{z}
& \left(2n_{C}+1\right)\mathbbm{1}_{2} & 2m_{CD}\,\sigma_{z}\\[4pt]
2m_{AD}\,\sigma_{z} & 2\kappa_{BD}\,\mathbbm{1}_{2}
& 2m_{CD}\,\sigma_{z} & \left(2n_{D}+1\right)\mathbbm{1}_{2}
\end{pmatrix},
\label{eq:covmatrix}
\end{equation}
where $\sigma_{z}=\mathrm{diag}(1,-1)$ and the entries are read off from Eqs. \eqref{eq:appn} and \eqref{eq:appcorr}. The purity of
$|0\rangle_{G}$ requires $\det\sigma=1$, which we have verified.

\subsection{General two-mode negativity}

The reduced two-mode states are obtained by deleting the appropriate rows and columns of
Eq. \eqref{eq:covmatrix}. For a reduced state
\begin{equation}
\sigma_{12}=\begin{pmatrix}
\left(2n_{1}+1\right)\mathbbm{1}_{2} & 2m\,\sigma_{z}\\
2m\,\sigma_{z} & \left(2n_{2}+1\right)\mathbbm{1}_{2}
\end{pmatrix},
\label{eq:sigma12}
\end{equation}
the partial-transpose symplectic eigenvalue is
\begin{equation}
\tilde{\nu}_{-}^{2}=\frac{1}{2}\left[
\tilde{\Sigma}-\sqrt{\tilde{\Sigma}^{2}-4\det\sigma_{12}}\right],
\qquad
\tilde{\Sigma}=\left(2n_{1}+1\right)^{2}+\left(2n_{2}+1\right)^{2}+8m^{2},
\label{eq:nugeneral}
\end{equation}
with
$\det\sigma_{12}=\left[\left(2n_{1}+1\right)\left(2n_{2}+1\right)-4m^{2}\right]^{2}$,
and $E_{\mathcal{N}}=\max\{0,-\log_{2}\tilde{\nu}_{-}\}$. For $n_{1}=n_{2}=n$ this reduces
to Eq. \eqref{eq:ENdef}. Pairs correlated only through the number-type entries
$\kappa_{AC},\kappa_{BD}$ (with vanishing $m$) are passively correlated and always
separable.

\section{The $\omega>0$ sector}
\label{app:omegapos}

\subsection{Charge fractions and local transformation}

For $\omega>0$ both $\mathcal{U}_{\mathrm{in},R}$ and $\mathcal{V}_{\mathrm{in},R}$ carry
charge $+1$ and are therefore both particle modes, the local in-out transformation is a
passive rotation and creates no pairs. From Eqs. \eqref{eq:q1}-\eqref{eq:q2} with
$\nu_{\omega}=+\nu$,
\begin{equation}
q_{2}=\frac{s(X-1)}{1+sX},
\qquad
q_{1}=\frac{1+s}{1+sX},
\qquad
q_{1}+q_{2}=1 .
\label{eq:qpos}
\end{equation}
The local relation reads
$\mathcal{U}^{\omega}_{\mathrm{out},R}
=\alpha^{R}_{\mathcal{U}\mathcal{U}}\mathcal{U}^{\omega}_{\mathrm{in},R}
+\alpha^{R}_{\mathcal{U}\mathcal{V}}\mathcal{V}^{\omega}_{\mathrm{in},R}$ with
$|\alpha^{R}_{\mathcal{U}\mathcal{U}}|^{2}=q_{2}$,
$|\alpha^{R}_{\mathcal{U}\mathcal{V}}|^{2}=q_{1}$, i.e.\ a beam splitter of reflectivity
$q_{2}$ acting on the pair $(\mathcal{U}_{R},\mathcal{V}_{R})$. Consequently
\begin{equation}
|0\rangle^{\mathrm{in}}_{R}=|0\rangle^{\mathrm{out}}_{R}
\qquad (\omega>0),
\label{eq:trivialvac}
\end{equation}
and there is no Schwinger squeezing in this sector.

\subsection{Global modes}

For $\omega>0$, the modes $\mathcal{U}_{\mathrm{in},R}$ and
$\mathcal{V}_{\mathrm{in},L}$ have Klein-Gordon norms $+1$ and
$-1$, respectively. They form the mixed pair in the Unruh
construction. The global state considered here is the Minkowski in-vacuum defined by the in-Unruh annihilation operators rather than the
Minkowski out-vacuum or the local Rindler in-vacuum.

To determine the mixing angle explicitly, we define
$\nu=\omega/a>0$ and introduce
\begin{equation}
\mathcal K_+^2
=e^{-\pi\Delta_n/2}
\frac{\cosh[\pi(\nu-\Delta_n/2)]}{\sinh(\pi\nu)},
\qquad
\mathcal R_+
=\frac{\Gamma[\tfrac12+i(\Delta_n/2-\nu)]}
{\Gamma(\tfrac12-i\Delta_n/2)}.
\label{eq:positive_mixing_prefactors}
\end{equation}
These quantities enter the coefficients of the global
mode transformation and determine the corresponding
mixing parameter. The global vacuum considered here
is defined by the annihilation operators associated
with the global in-modes and should be distinguished
from the local Rindler in-vacuum and the outgoing vacuum.
By restricting the particle in-Unruh mode given in \cite{Gabriel:1999yz} to
$R\cup L$ and making the replacement $m^2\to M_n^2$ gives,
up to an overall phase,
\begin{equation}
\Phi_{\omega}^{(p,\mathrm{in})}\big|_{R\cup L}
=\mathcal K_+
\left[
 i e^{\pi\nu/2}\mathcal R_+
 \mathcal U_{\mathrm{in},R}^{\omega}
 +\mathcal V_{\mathrm{in},L}^{\omega}
\right].
\label{eq:positive_global_mode_explicit}
\end{equation}
Here the transverse labels $(n,k_z)$ are suppressed. The associated
antiparticle in-Unruh mode supplies the conjugate canonical
transformation.
Now using $|\Gamma(\tfrac12+iy)|^2=\pi/\cosh(\pi y)$ gives
\begin{equation}
|\mathcal R_+|^2
=\frac{\cosh(\pi\Delta_n/2)}
{\cosh[\pi(\nu-\Delta_n/2)]}.
\label{eq:positive_gamma_modulus}
\end{equation}
Let $\mathcal A_+$ and $\mathcal B_+$ denote the coefficients
of the positive- and negative-norm local modes in Eq. \eqref{eq:positive_global_mode_explicit}. Their moduli are given by
\begin{align}
|\mathcal A_+|^2
&=\mathcal K_+^2 e^{\pi\nu}|\mathcal R_+|^2
=\frac{X(1+s)}{X-1},
\label{eq:positive_alpha_modulus}\\
|\mathcal B_+|^2
&=\mathcal K_+^2
=\frac{1+sX}{X-1}.
\label{eq:positive_beta_modulus}
\end{align}
Thus $|\mathcal A_+|^2-|\mathcal B_+|^2=1$, confirming the
unit Klein-Gordon norm and the bosonic canonical relation.
With $|\mathcal A_+|=\cosh r_U^{(+)}$ and
$|\mathcal B_+|=\sinh r_U^{(+)}$, it follows that
\begin{equation}
\tanh^2 r_U^{(+)}
=\frac{|\mathcal B_+|^2}{|\mathcal A_+|^2}
=\frac{e^{-\pi\nu}\cosh[\pi(\nu-\Delta_n/2)]}
{\cosh(\pi\Delta_n/2)}
=\frac{1+sX}{X(1+s)},
\qquad
\sinh^2 r_U^{(+)}=\frac{1+sX}{X-1}.
\label{eq:Zpos}
\end{equation}
This establishes the mixing angle directly from the normalised
in-Unruh coefficients. Norm conservation alone would not fix
its value without the relative coefficient supplied by the
global mode construction.

As $s\to0$, these expressions reduce to $e^{-2\pi\nu}$ and
$(e^{2\pi\nu}-1)^{-1}$, respectively. At fixed $X>1$, both
increase with $s$, in particular,
\begin{equation}
\frac{\partial}{\partial s}
\left[\frac{1+sX}{X(1+s)}\right]
=\frac{X-1}{X(1+s)^2}>0.
\label{eq:monotonepos}
\end{equation}
The difference between Eqs. \eqref{eq:Z} and \eqref{eq:Zpos}
reflects the frequency-dependent particle and antiparticle
content of the Minkowski in-vacuum in the wedge basis.
The charge polarisation concerns the complete particle and
antiparticle populations, including both frequency sectors,
rather than a single mode-pair contribution.

\subsection{Number densities}

Since the local in- and out-vacua coincide in the
positive-frequency sector, as shown in
Eq.~\eqref{eq:trivialvac}, the outgoing particle
number densities can be obtained from the global
Unruh occupation $\sinh^2 r_U^{(+)}$ through the
local beam-splitter transformation. We obtain
\begin{align}
n_{\mathcal{U}^{\mathrm{out}}_{R}}
&=q_2\,\sinh^2 r_U^{(+)}
=\frac{s(X-1)}{1+sX}
\frac{1+sX}{X-1}
=e^{-\pi\Delta_n},
\label{eq:nUpos}\\[4pt]
n_{\mathcal{V}^{\mathrm{out}}_{R}}
&=q_1\,\sinh^2 r_U^{(+)}
=\frac{1+s}{1+sX}
\frac{1+sX}{X-1}
=\frac{1+e^{-\pi\Delta_n}}
{e^{2\pi|\omega|/a}-1}.
\label{eq:nVpos}
\end{align}
The corresponding left-wedge densities satisfy
$n_{\mathcal{V}^{\mathrm{out}}_{L}}
=n_{\mathcal{U}^{\mathrm{out}}_{R}}$
and
$n_{\mathcal{U}^{\mathrm{out}}_{L}}
=n_{\mathcal{V}^{\mathrm{out}}_{R}}$.

Equation~\eqref{eq:nUpos} agrees with
Eq.~\eqref{eq:nU}, showing that the Schwinger
mean occupation $e^{-\pi\Delta_n}$ is the same
in both frequency sectors. However, the remaining
outgoing-mode occupations differ. Comparing
Eqs.~\eqref{eq:nVpos} and \eqref{eq:nV}, we find
that the negative-frequency occupation exceeds
the positive-frequency result by
$e^{-\pi\Delta_n}$ at the same value of
$|\omega|/a$.

\subsection{Negativity}

The local in-out transformations in the positive-frequency sector
are passive beam splitters. They mix each member of the entangled
in-mode pair with an initially unoccupied mode within the same
wedge. As local unitaries, they preserve the entanglement across
the complete $R|L$ partition. A selected pair of out-modes,
however, is obtained by tracing over the other two outputs.
For that reduced pair, the transformation is equivalent to
independent attenuation channels with vacuum environments,
its correlations are attenuated and the state is generally mixed.
Writing $N^{(+)}=\sinh^2r_U^{(+)}$ and
$M^{(+)}=\sqrt{N^{(+)}(1+N^{(+)})}$, an output mode with
intensity weight $f\in\{q_1,q_2\}$ on $R$, paired with a mode
of weight $g$ on $L$, has the moments

\begin{equation}
n_{1}=f\,N^{(+)},
\qquad
n_{2}=g\,N^{(+)},
\qquad
|m|=\sqrt{fg}\;M^{(+)} .
\label{eq:posmoments}
\end{equation}

For the two symmetric pairings, $f=g$, the reduced covariance
matrix has identical local blocks
$[1+2fN^{(+)}]\mathbbm 1_2$ and correlation block
$2fM^{(+)}\sigma_z$. Its two symplectic eigenvalues are
\begin{equation}
\mu_f=\sqrt{1+4f(1-f)N^{(+)}}.
\label{eq:positive_pair_mixedness}
\end{equation}
Thus, for $N^{(+)}>0$ and $0<f<1$, the selected pair is mixed,
even though the complete four-mode state is pure. The
partial-transpose criterion in Eq. \eqref{eq:ENdef} gives

\begin{equation}
\tilde{\nu}_{-}=1+2f\left(N^{(+)}-M^{(+)}\right)<1
\qquad\text{since}\qquad
N^{(+)}<\sqrt{N^{(+)}\left(1+N^{(+)}\right)},
\label{eq:posent}
\end{equation}

The strict inequality holds for $f>0$ and $N^{(+)}>0$.
At $f=0$ or $N^{(+)}=0$, the logarithmic negativity vanishes.
Equivalently,
\begin{equation}
\widetilde\nu_-
=(1-f)+f e^{-2r_U^{(+)}},
\qquad
0\leq E_{\mathcal N}^{(f)}
\leq\frac{2r_U^{(+)}}{\ln2}.
\label{eq:positive_pair_attenuation}
\end{equation}
For finite nonzero $r_U^{(+)}$, the upper bound is reached only
at $f=1$, when the complete entangled pair is retained.
For $0<f<1$, tracing over the other outputs reduces the
pairwise logarithmic negativity. Explicitly,

\begin{align}
E_{\mathcal{N}}\!\left(\mathcal{U}^{\mathrm{out}}_{R}{:}\,
\mathcal{V}^{\mathrm{out}}_{L}\right)
&=-\log_{2}\left[1+2s\left(1-\sqrt{\frac{X(1+s)}{1+sX}}\right)\right],
\label{eq:ENposA}\\[4pt]
E_{\mathcal{N}}\!\left(\mathcal{V}^{\mathrm{out}}_{R}{:}\,
\mathcal{U}^{\mathrm{out}}_{L}\right)
&=-\log_{2}\left[\frac{X-1+2(1+s)
-2(1+s)^{3/2}\sqrt{X/(1+sX)}}{X-1}\right].
\label{eq:ENposB}
\end{align}
Equation \eqref{eq:ENposB} reduces to $\log_{2}\coth(\pi\nu/2)$ at $s=0$, while
Eq. \eqref{eq:ENposA} vanishes there, both as expected. The asymmetric pairings require the
general formula Eq. \eqref{eq:nugeneral}.

For finite $X>1$ and $s>0$, both symmetric
cross-wedge particle--antiparticle pairs
remain entangled in the positive-frequency
sector. However, neither pair generally
contains the complete entanglement between
the right and left Rindler wedges.
This follows from the passive nature of
the local in--out transformations, which
redistribute the correlations among the
outgoing modes without changing the
entanglement across the complete wedge
partition.

The situation is different for $\omega<0$,
where the local transformation corresponds
to two-mode squeezing. In that sector,
the reduced pair $A:D$ remains separable,
while $B:C$ is entangled only when
the condition in Eq.~\eqref{eq:thresholdexplicit}
is satisfied. These results show that the
entanglement of the selected outgoing pairs
depends on the frequency sector and the
corresponding Bogoliubov transformation.

\subsection{Total $R|L$ entanglement}

As in Sec. \ref{sec:negativity}, the beam splitter is a local unitary and
\begin{equation}
E_{\mathcal{N}}(R|L)\Big|_{\omega>0}
=\log_{2}\frac{1+\tanh r_{U}^{(+)}}{1-\tanh r_{U}^{(+)}}
=\log_{2}\left[\frac{\sqrt{X(1+s)}+\sqrt{1+sX}}{\sqrt{X(1+s)}-\sqrt{1+sX}}\right],
\label{eq:ENtotalpos}
\end{equation}
again monotonically increasing with the electric field.

\bibliography{Schwinger_references_format_corrected}

\end{document}